\documentclass{IEEEtran}
\usepackage{amsmath}
\usepackage{amssymb}
\usepackage{amsfonts}
\usepackage{graphicx}
\usepackage{tudacolors}
\usepackage{tikz}
\usepackage{pgfplots}
\usepackage{pgfplotstable} 
\usepackage{booktabs}
\usepackage{hyperref}
\usepackage{makecell}
\makeatletter
\let\MYcaption\@makecaption
\makeatother

\usepackage[font=footnotesize]{subcaption}

\makeatletter
\let\@makecaption\MYcaption
\makeatother
\usepackage{siunitx}
\usepackage{url}

\usetikzlibrary{shapes.geometric, calc, shapes, positioning, backgrounds}
\tikzstyle{startstop} =[rectangle, rounded corners, minimum width=.08\textwidth, minimum height=2em, text centered, draw=black, fill=white]
\tikzstyle{io} =[trapezium, trapezium left angle=70, trapezium right angle = 110,minimum width=.08\textwidth, minimum height=2em, text centered, draw=black, fill=white]
\tikzstyle{process}=[rectangle, minimum width=.08\textwidth, minimum height=2em, text centered, draw=black, fill=white]
\tikzstyle{decision}=[diamond,  minimum width=.08\textwidth, minimum height=2em,  text centered, draw=black, fill=white, aspect=2]
\tikzstyle{arrow}=[thick, ->, >=stealth]

\def\myspace{\mspace{2mu}}

\usepackage{eso-pic}
\AddToShipoutPicture*{\footnotesize\sffamily\raisebox{0.75cm}{\hspace{1.65cm}\fbox{\parbox{\textwidth}{\copyright 2022 IEEE.  Personal use of this material is permitted.  Permission from IEEE must be obtained for all other uses, in any current or future media, including reprinting/republishing this material for advertising or promotional purposes, creating new collective works, for resale or redistribution to servers or lists, or reuse of any copyrighted component of this work in other works.}}}}

\begin{document}
\title{Mode Recognition by Shape Morphing for Maxwell's Eigenvalue Problem}
\author{%
	Anna~Ziegler,~\IEEEmembership{~}
	Niklas~Georg,~\IEEEmembership{~}
	Wolfgang~Ackermann,~\IEEEmembership{~}
	Sebastian~Schöps~\IEEEmembership{~}
	\thanks{Manuscript received xxxx yy, 2022; revised xxxxx y, 2022; accepted xxxxx y, 2022. Date of publication xxxxx y, 2022; date of current version xxxxx y, 2022.
	All authors are with the institute for Accelerator Science and Electromagnetic Fields (TEMF) at TU Darmstadt, Darmstadt, Germany (e-mails:  anna.ziegler@tu-darmstadt.de, wolfgang.ackermann@tu-darmstadt.de, sebastian.schoeps@tu-darmstadt.de). 
	A.~Ziegler, N.~Georg and S.~Schöps are with the Graduate School of Computational Engineering at TU Darmstadt, Darmstadt, Germany.
	The authors thank Herbert De Gersem, Jacopo Corno and Mario Mally for the support and the many fruitful discussions.
	This work is supported by the Graduate School CE within the Centre for Computational Engineering at Technische Universität Darmstadt and by the Federal Ministry of Education and Research (BMBF) and the state of Hesse as part of the NHR Program. Corresponding author: A. Ziegler.\\
	Color versions of one or more of the figures in this article are available online at \url{http://ieeexplore.ieee.org}.\\
Digital Object Identifier 10.1109/TAP.2022.xxxxxxx}
}

\maketitle

\begin{abstract}
In electrical engineering, for example during the design of superconducting radio-frequency cavities, eigenmodes must be identified based on their field patterns. 
This allows to understand the working principle, optimize the performance of a device and distinguish desired from parasitic modes. 
For cavities with simple shapes, the eigenmodes are easily classified according to the number of nodes and antinodes in each direction as is obvious from analytical formulae. 
For cavities with complicated shapes, the eigenmodes are determined numerically.
Thereby, the classification is cumbersome, if not impossible.
In this paper, we propose a new recognition method by morphing the cavity geometry to a pillbox and tracking its eigenmodes during the deformation.
\end{abstract}

\begin{IEEEkeywords}
Electromagnetic Wave Equation, Resonant Cavities, Eigenmodes, Isogeometric Analysis
\end{IEEEkeywords}
%------------------------------------------------------------------------
\section{Introduction}
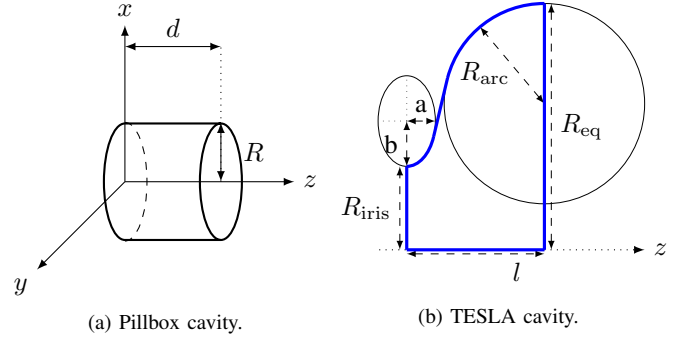
\begin{figure}
	\begin{subfigure}{0.5\linewidth}
		\centering	
        \def\height{0.075\textheight}
\def\width{0.35\textwidth}

\begin{tikzpicture}

    %the cylinder
    \node[cylinder,draw=black,thick,aspect=0.7,
        minimum height=\height,minimum width=\width,
        shape border rotate=0,
        cylinder uses custom fill,
        ]
        (A) at (0,0) {\phantom{\Huge A}};
    \draw[dashed]
        let \p1 = ($ (A.after bottom) - (A.before bottom) $),
            \n1 = {0.5*veclen(\x1,\y1)},
            \p2 = ($ (A.bottom) - (A.after bottom)!.5!(A.before bottom) $),
            \n2 = {veclen(\x2,\y2)}
     in
        (A.before bottom) arc [start angle=270, end angle=450,
        x radius=\n2, y radius=\n1];

    %the axes    
    \path let \p3 = (A.before bottom) in node (origin)  at (\x3,0) {};
    \path let \p4 = (A.before top) in node (top)  at (\x4,0) {};
    \draw[-latex] (origin.center) -- (\height,0) node [pos=1, right] {$z$};
    \node (xaxis) [above = \height of origin] {};
    \draw[-latex] (origin.center) -- (xaxis.center) node [pos=1, above] {$x$};
    \node (yaxis) [below left = 0.71*\height of origin] {};
    \draw[-latex] (origin.center) -- (yaxis.center) node [pos=1, below left] {$y$};
    
    %the labels
    \node (pt1) [above = \width of origin] {};
    \node (pt2) [above = \width of top] {};
    \draw[latex-latex] (pt1.center) -- (pt2.center) node [midway, above] {$d$};
    \draw[dotted] (top.center) -- (pt2.center);
    \draw[latex-latex] (top.center) -- (A.before top) node [midway, right, xshift = 0.5em] {$R$};

\end{tikzpicture}
		\vspace*{-1.5em}
		\caption{Pillbox cavity.}
		\label{fig:pillbox_shape}
	\end{subfigure}%
	\begin{subfigure}{0.5\linewidth}
		\centering	
		\includegraphics{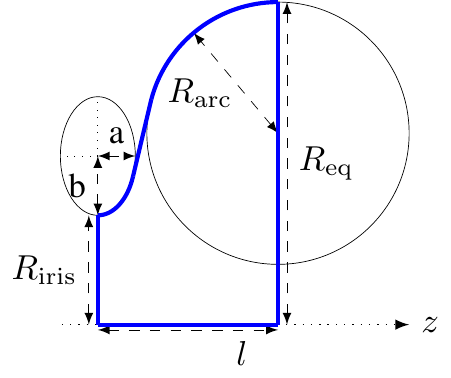}
% 		\vspace*{-2em}
		\caption{TESLA cavity.}
		\label{fig:tesla_shape}
	\end{subfigure}%
	\caption{Geometrical shape of pillbox and TESLA cavity.}
\end{figure}

%------------------------------------------------------------------------
\IEEEPARstart{C}{avity} resonators are structures in which a standing wave, with different modes and frequencies, is formed by resonance. One of the most well-known examples of a lossless resonator is the pillbox cavity, see Fig.~\ref{fig:pillbox_shape}. 
Modern resonators have more complex shapes, e.g. each cell of the superconducting radiofrequency TESLA cavity \cite{Aune_2000aa} that is used for particle acceleration has been constructed by several ellipses quantified in terms of six geometry parameters (lengths and radii), see Fig.~\ref{fig:tesla_shape}. Since the shape is a major design choice, computer simulation, shape optimization and uncertainty quantification are part of nowadays design processes \cite{Shemelin_2003aa,Georg_2019aa,Corno_2020aa}.

Originally, low order discretization methods, e.g. based on the finite difference or finite integration method, were used to numerically approximate the underlying Maxwell eigenvalue problem, see e.g. \cite{Halbach_1976aa,van-Rienen_1985aa,Weiland_1985aa}. Nowadays, finite elements (FE) with higher-order basis functions and curved elements are state of the art \cite{Ainsworth_2003ab,Zaglmayr_2006aa}. Recently, a new finite-element variant called isogeometric analysis (IGA) was proposed by Cottrell et al. \cite{Cottrell_2009aa} and firstly applied to the simulation of cavities, by Corno et al. \cite{Corno_2016aa}. IGA uses the same spline-based functions from computer-aided-design (CAD) to describe solution and geometry such that no geometry-related modeling error is introduced. This is particularly useful when dealing with problems that are sensitive with respect to their geometric shape.

\begin{figure}
	\centering
	\begin{subfigure}{0.3\linewidth}
		\centering	
		\includegraphics[trim = 43em 13em 43em 13em, clip, width=.9\textwidth]{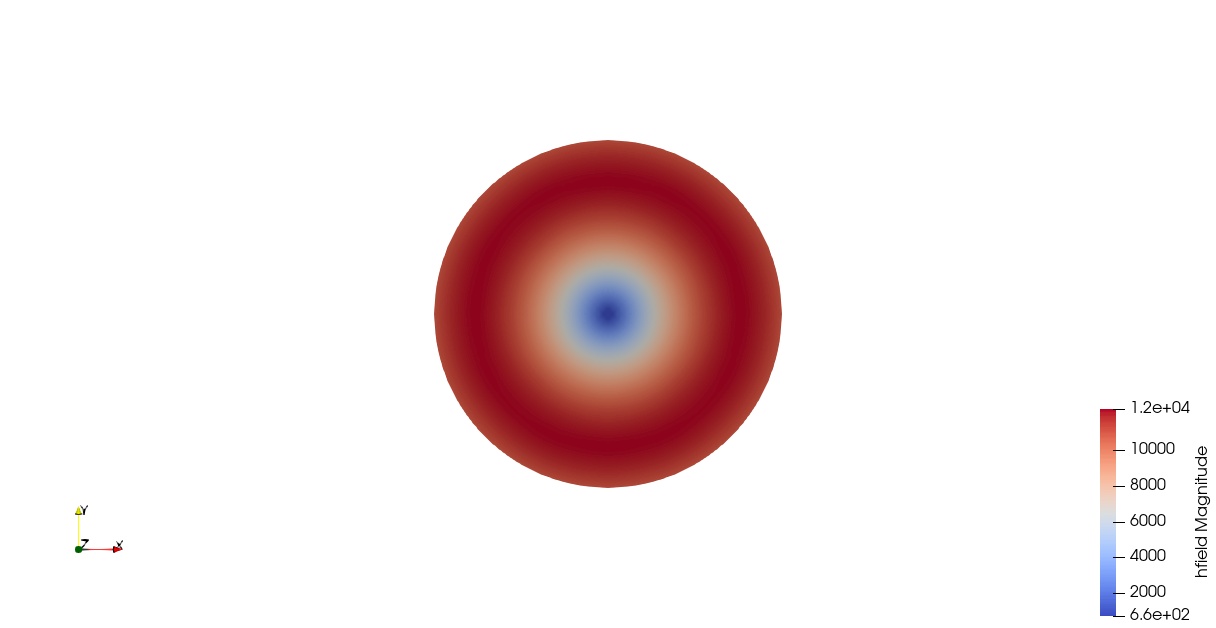}
		\caption{$\mathrm{TM}0\myspace1\myspace p$-mode}
	\end{subfigure}%
	\begin{subfigure}{0.3\linewidth}
		\centering	
		\includegraphics[trim = 43em 13em 43em 13em, clip, width=.9\textwidth]{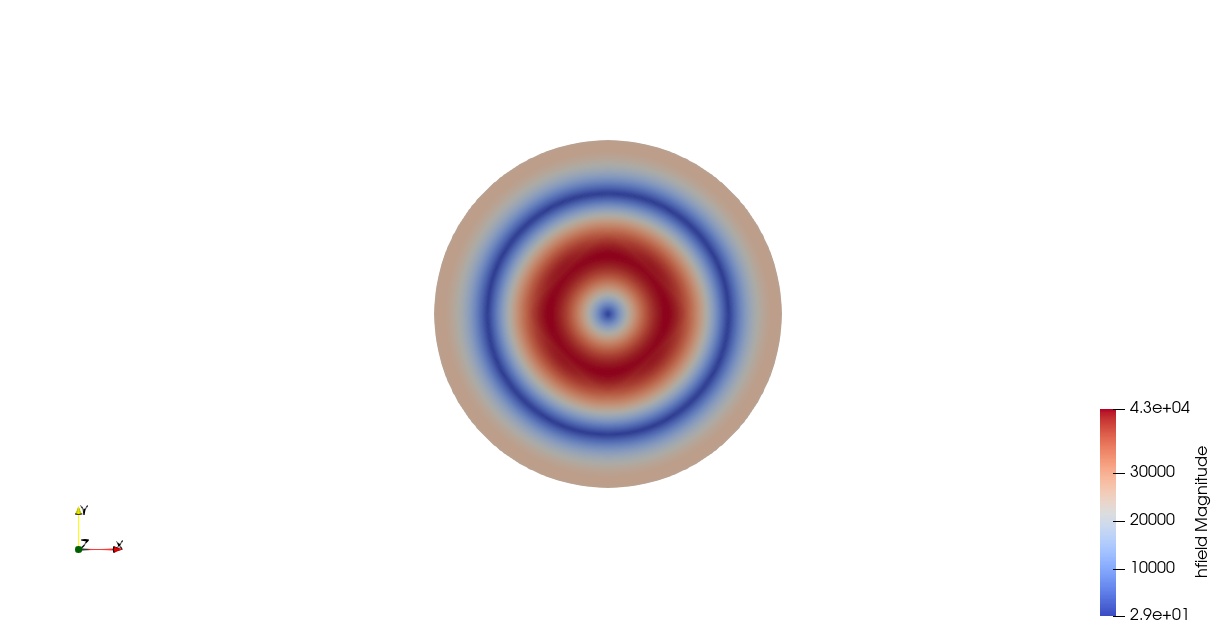}
		\caption{$\mathrm{TM}0\myspace2\myspace p$-mode}
	\end{subfigure}%
	\begin{subfigure}{0.3\linewidth}
		\centering	
		\includegraphics[trim = 43em 13em 43em 13em, clip, width=.9\textwidth]{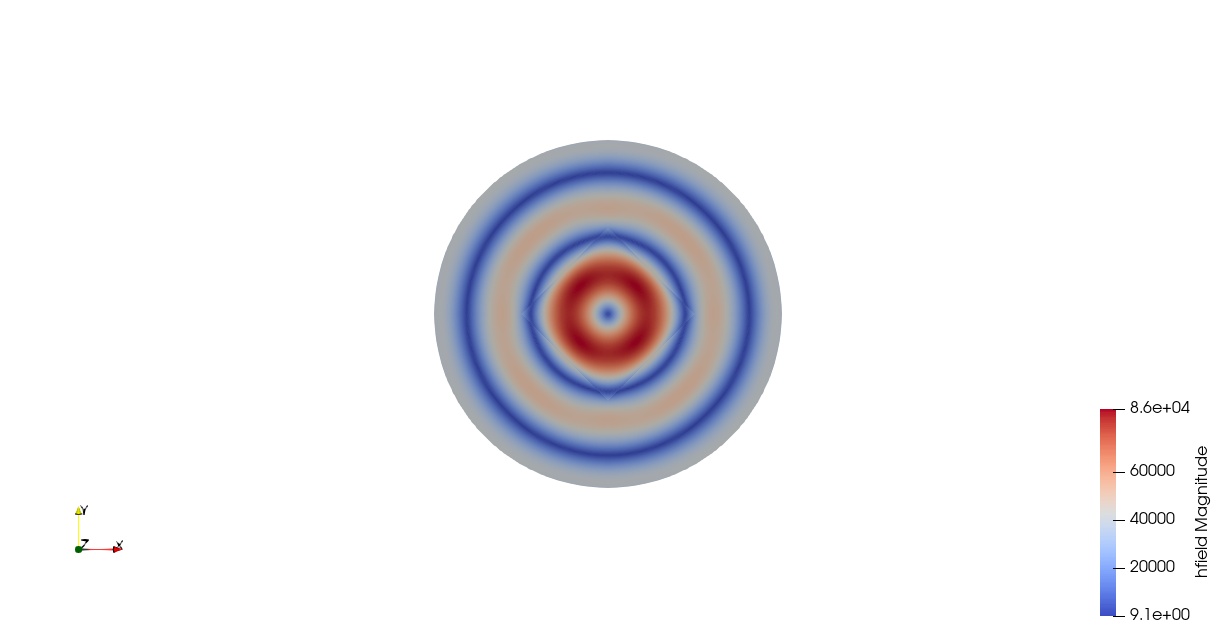}
		\caption{$\mathrm{TM}0\myspace3\myspace p$-mode}
	\end{subfigure}%
	\caption{\label{fig:pillbox_modes}Exemplary mode patterns in pillbox cavity, where $p$ is an arbitrary integer value denoting the number of half-waves in longitudinal direction (not identifiable in 2D).}
\end{figure}

In accelerator science, often many eigenmodes and eigenfrequencies must be computed by solving Maxwell's eigenvalue problem, for example Ackermann et al. and Wanzenberg investigate the first 194 modes and 196 modes, respectively, of the TESLA cavity for DESY in \cite{Ackermann_2015aa, Wanzenberg_2001aa}. They identify the modes and the corresponding passbands entirely manually or based on the azimuthal classification since a reliable automatic mode recognition is not available until today. Existing approaches may identify modes wrongly or are not applicable in the case of multiple cells or high frequencies, see \cite{Brackebusch_2014aa}. To this end, we propose here a new method that is based on shape morphing and eigenvalue tracking: we gradually deform the geometry of a complex shaped cavity to a simple one, e.g., the cylindrical pillbox. This allows us to uniquely identify each eigenpair of the complex shaped cavity with one of the pillbox, where modes, frequencies and their categorization are known in closed form, e.g.,
\begin{align}
	\omega_{\mathrm{TM}, mnp} 
	&= 
	\frac{1}{\sqrt{\mu \varepsilon}} \sqrt{\frac{x^2_{mn}}{R^2} + \frac{p^2 \pi ^2}{d^2}}
	\label{eq:resonanceFrequenciesTM}
	\\
    \intertext{resp.}
    \omega_{\mathrm{TE},mnp} 
	&=
	\frac{1}{\sqrt{\mu \varepsilon}} \sqrt{\frac{x'^2_{mn}}{R^2} + \frac{p^2 \pi ^2}{d^2}},
	{\label{eq:resonanceFrequenciesTE}}
\end{align}
where $\mu$ denotes the permeability, $\varepsilon$ the permittivity, $R$ indicates the radius and $d$ the length of the pillbox. The expression~$x_{mn}$ indicates the $n$th root of the Bessel function $J_m(x) = 0$ and $x'_{mn}$ the $n$th root of $J'_m(x) = 0$.
Integers~$m$,~$n$ and~$p$ take on the values $m = 0, 1, 2,\ldots$, $n = 1, 2, 3,\ldots$ and $p = 0, 1, 2,\ldots$, where the lowest transverse magnetic ($\mathrm{TM}$) mode has $m = p = 0$ and $n = 1$  (hence is classified as $\mathrm{TM}{0\myspace1\myspace0}$). The lowest transverse electric ($\mathrm{TE}$) mode has $m = n = p = 1$ (hence $\mathrm{TE}{1\myspace1\myspace1}$). Details can be found in \cite[Section 8.7]{Jackson_1998aa} and the patterns of the first three TM modes are visualized in Fig.~\ref{fig:pillbox_modes}. 

The paper is structured as follows.
In Section~\ref{sec:domain} we describe the geometry and  the parametrization of the computational domain.
Subsequently, we state the Maxwell eigenvalue problem and introduce its isogeometric discretization in Section~\ref{sec:maxwell}.
In Section~\ref{sec:classification} we present our mode classification algorithm and the possibility to use it both for identifying and for exploring modes in a complex cavity geometry.
We demonstrate the application of our algorithm for conveying the well-known mode nomenclature of the cylindrical pillbox cavity to more complex cavities in Section~\ref{sec:application}.
Furthermore, we automatically recognize modes of the TESLA cavity by morphing its shape to the pillbox cavity and discuss the results.
We conclude our work in Section~\ref{sec:conclusion}.

%------------------------------------------------------------------------
\section{Domain Parametrization} \label{sec:domain}
\begin{figure}%
    \begin{subfigure}{0.5\linewidth}
		\centering	
		\includegraphics[width=\textwidth]{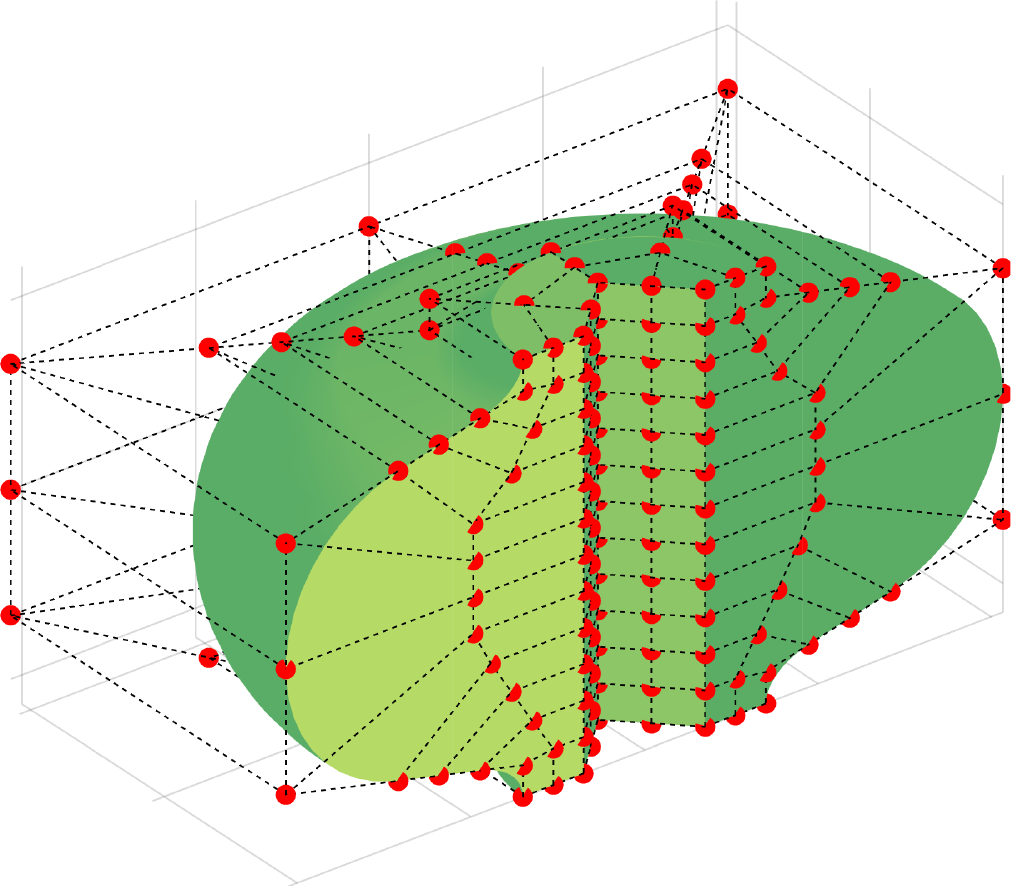}
		\caption{Mesh at $t=0$}
	\end{subfigure}%
	\begin{subfigure}{0.5\linewidth}
		\centering	
		\includegraphics[width=\textwidth]{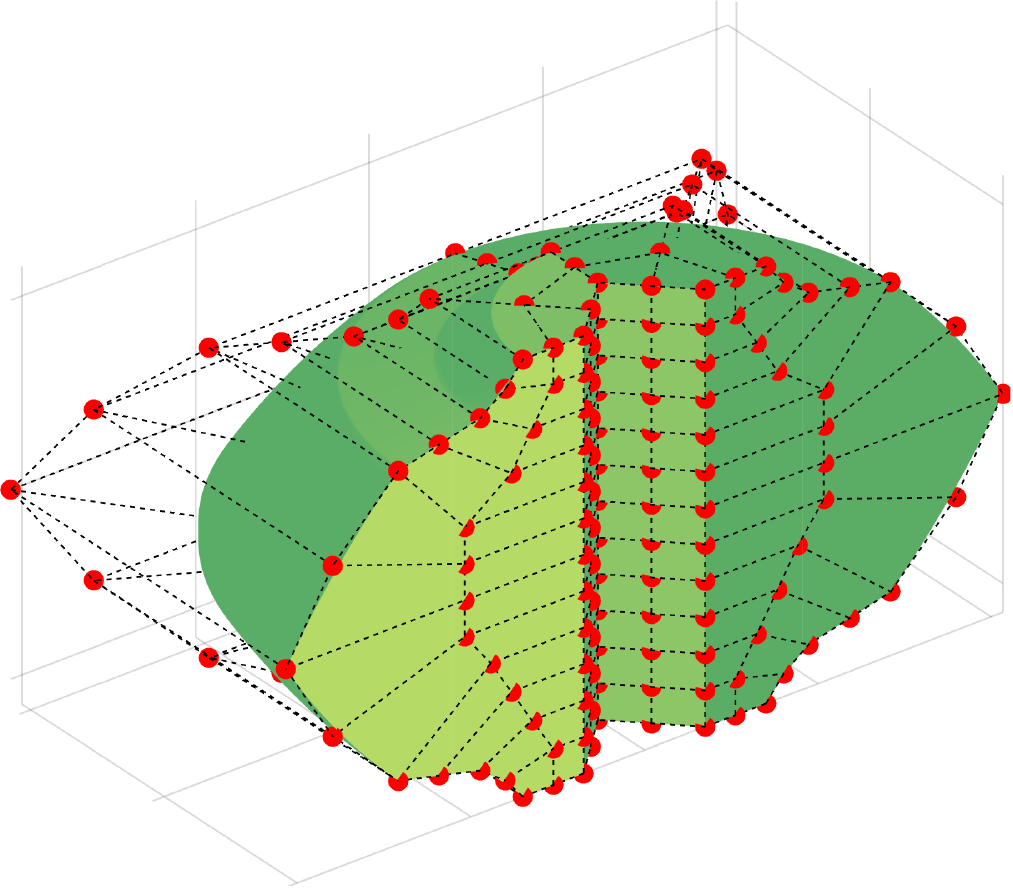}
		\caption{Mesh at $t=0.4$}
	\end{subfigure}
	\begin{subfigure}{0.5\linewidth}
		\centering	
		\includegraphics[trim = 2em 0em 2em 0em, clip, width=\textwidth]{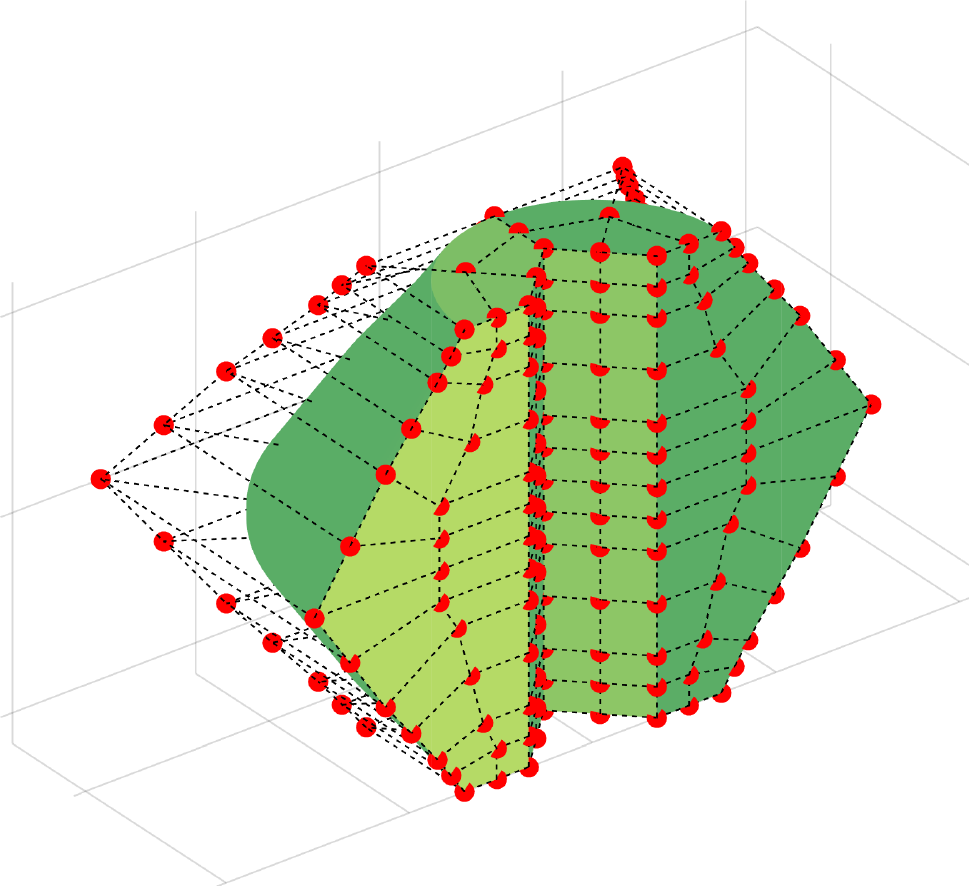}
		\caption{Mesh at $t=0.7$}
	\end{subfigure}%
	\begin{subfigure}{0.5\linewidth}
		\centering	
		\includegraphics[trim = 10em 3em 6em 2em, clip, width=\textwidth]{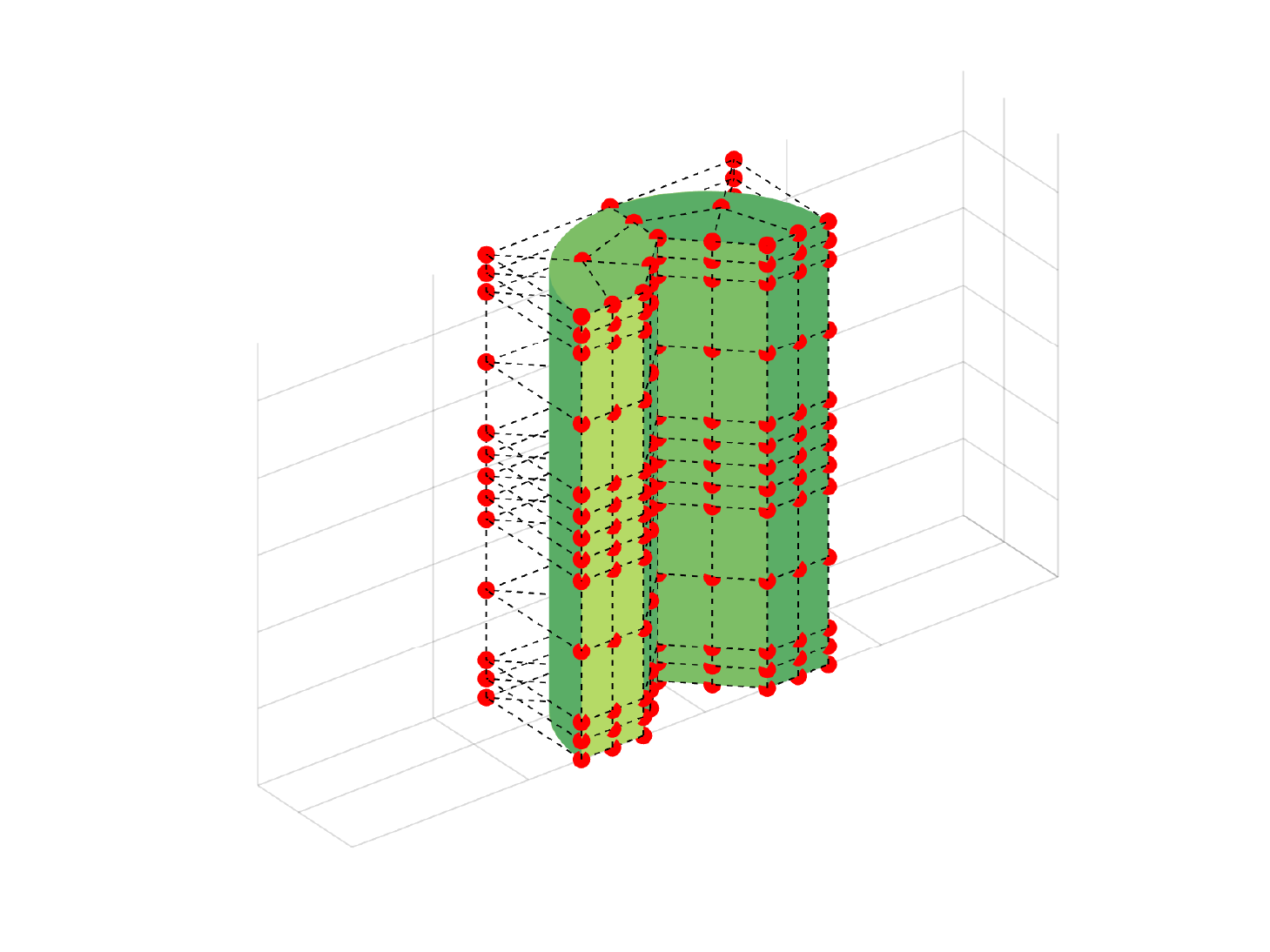}
		\caption{Mesh at $t=1$}
	\end{subfigure}%
    \caption{Illustration of a mesh deformation which transforms the shape of a TESLA cell to a pillbox shape. The control points are illustrated in red. The full cavity is composed of 5 patches, see \cite[Section 3.6]{Corno_2017ad} for details.}
    \label{fig:IGAmeshDeformation}
\end{figure}

%------------------------------------------------------------------------
Due to the relevance of the cavity shape, we start with a brief summary of
geometry description in terms of CAD \cite{Cohen_2001aa}. The physical domain 
$\Omega$ is represented as a mapping
\begin{equation}\label{eq:iga-map}
	\mathbf{G}:\hat{\Omega} \rightarrow \Omega
\end{equation}
from a reference domain $\hat{\Omega}$. In CAD, this reference domain
is commonly the unit cube $\hat{\Omega}=[0,1]^3$. If a shape is too
complex to be represented by a single mapping, one uses several mappings, 
also called patches, to represent the overall geometry. Each patch is
given by a B-spline or Non-Uniform Rational B-spline (NURBS) curve. We discuss here, for simplicity, 
only the case of a trivariante B-spline mapping, i.e.,
\begin{equation}
	\label{eq:volumetric_bspline}
	\mathbf{G}(\mathbf{\hat{x}}) = \sum_{i=1}^{n_1} \sum_{j=1}^{n_2} \sum_{k=1}^{n_3}  \mathbf{P}_{i,j,k} B_{i,p_1} B_{j,p_2} B_{k,p_3}\
\end{equation}
with control points $\mathbf{P}_{i,j,k}$ that form the so-called control mesh. The case of NURBS is similar but requires the introduction of an additional weighting. Each B-spline basis functions $\{ B_{i,p} \}_{i=1}^{N}$ stems from 
a one-dimensional B-spline space $S^{p}_{\alpha}$ of degree $p$ and regularity $\alpha$ \cite[Chapter 7]{Cohen_2001aa}. They are constructed from a knot vector $\boldsymbol{\Xi} = (\xi_1, \xi_2, \dots, \xi_{n}) \in [0,1]^n$, $\xi_1 \leq \xi_2 \leq \dots \leq \xi_n$ using the Cox-de Boor algorithm \cite{de-Boor_2001aa}
\begin{align*}
    B_{i,0}(\xi) &= \begin{cases}
        1 \quad \mathrm{if} \quad \xi_i \leq \xi < \xi_{i+1}\\
        0 \quad \mathrm{otherwise}
    \end{cases}\\
    B_{i,p}(\xi) &= \frac{\xi - \xi_i}{\xi_{i+p} - \xi_i} B_{i,p-1}(\xi) + \frac{\xi_{i+p+1} - \xi}{\xi_{i+p+1} - \xi_{i+1}} B_{i+1,p-1}(\xi)\,.
\end{align*}
Varying the control points allows us to gradually morph one shape to another. 
For this paper it is sufficient to consider linear transformations of the control points: let us assume we have two cavities represented by two mappings $\mathbf{G}^{(1)}$ and $\mathbf{G}^{(2)}$ with matching control meshes, i.e. $\mathbf{P}_{i,j,k}^{(1)}$ and $\mathbf{P}_{i,j,k}^{(2)}$ of compatible dimensions. Then, we morph one into the other by the convex combination
\begin{align*}
	\mathbf{P}_{i,j,k}(t) := (1-t)\mathbf{P}_{i,j,k}^{(1)} + t \mathbf{P}_{i,j,k}^{(2)}
\end{align*}
when varying $t\in[0,1]$, see Fig.~\ref{fig:IGAmeshDeformation}. This gives raise to a parametrized version of~\eqref{eq:volumetric_bspline}
\begin{equation}
	\label{eq:iga-map-t}
	\mathbf{G}_t(\mathbf{\hat{x}}) 
	=
	\sum_{i=1}^{n_1}
	\sum_{j=1}^{n_2}
	\sum_{k=1}^{n_3}
	\mathbf{P}_{i,j,k}(t) 
	B_{i,p_1} B_{j,p_2} B_{k,p_3}
\end{equation}
and as a result also to the parametrized computational domain~$\Omega_t$.

%------------------------------------------------------------------------
\section{Problem Statement} \label{sec:maxwell}
%------------------------------------------------------------------------
Let us formulate in this section the electromagnetic field problem and its discretization.
%------------------------------------------------------------------------
\subsection{Maxwell's Eigenvalue Problem}
%------------------------------------------------------------------------
Starting from Maxwell's equations on the domain $\Omega_t$, with homogeneous, lossless materials, perfect electric boundary conditions on $\partial\Omega_t$ and without sources, 
one derives Maxwell's eigenvalue problem: 
find wave numbers $k:=\omega \sqrt{\mu\varepsilon} \in \mathbb{R}^+$ and $\mathbf{E} \neq 0$ such that
\begin{equation}\label{eq:Maxwell-eig-disc}
\begin{aligned}
\textrm{curl}\,{\left(\textrm{curl}\,{\mathbf{E}}\right)} &= k^2 \mathbf{E} && \text{in }\Omega_t\\
\mathbf{E}\times\mathbf{n} &= 0 && \text{on }\partial\Omega_t.
\end{aligned}
\end{equation}
The weak formulation is obtained by testing and integration. The resulting problem reads: find $k\in \mathbb{R}^+$ and $\mathbf{E} \in H_0(\mathrm{curl};\Omega_t)$ such that
\begin{equation}
 \left(\textrm{curl}\,{\mathbf{E}},\textrm{curl}\,{\mathbf{v}}\right) 
  = k^2 \left(\mathbf{E},\mathbf{v}\right) \quad \forall \mathbf{v} \in H_0(\mathrm{curl};\Omega_t), 
\end{equation}
where we made use of the function space of square-integrable vector fields with square-integrable curl and vanishing trace $H_0(\mathrm{curl};\Omega_t)$, see \cite{Monk_2003aa}. Following the Ritz-Galerkin paradigm, we represent the unknown solution by basis functions $\mathbf{v}_j$ from the same space, i.e., 
\begin{equation}
	\mathbf{E} = \sum\nolimits_j e_{j} \mathbf{v}_{j}.
\end{equation}
Finally, we approximate the solution by considering only a finite-dimensional subspace $V_h \subset H_0(\mathrm{curl};\Omega_t)$ of dimension $N=\mathrm{dim}V_h$.
We obtain the generalized eigenvalue problem: find all $\mathbf{e}$ and $\lambda$ such that
\begin{equation}\label{eq:eigprob}
	\mathbf{K} \mathbf{e} = \lambda \mathbf{M} \mathbf{e},
\end{equation}
where $\lambda=k^2$, the parameter-dependent stiffness and mass matrices are expressed by
\begin{equation}
	\label{eq:matrix-def}
	\begin{aligned}
		\mathbf{K}_{i,j}(t) 
		&= 
		\int_{\Omega_t}\textrm{curl}\,\mathbf{v}_{j}\cdot\textrm{curl}\,\mathbf{v}_{i}\,\mathrm{d}\mathbf{x},
		\\
		\mathbf{M}_{i,j}(t)
		&=
		\int_{\Omega_t}\mathbf{v}_{j} \cdot\mathbf{v}_{i} \,\mathrm{d}\mathbf{x}
	\end{aligned}
\end{equation}
with $i,j=1,\ldots,N$. The parameter-dependence of the matrices is naturally inherited by the solutions of the eigenvalue problem
\begin{equation}
	\lambda=\lambda(t),
	\quad\text{and}\quad
	\mathbf{e}=\mathbf{e}(t)
\end{equation}
depending on the parameter $t$. 
Georg et al. \cite{Georg_2019aa} propose algebraic, simplified matrix mappings to replace \eqref{eq:matrix-def}, i.e.
\begin{subequations}
	\label{eq:homotopy}
	\begin{align}
		\mathbf{\tilde{K}}(t) &:=(1-t)\mathbf{K}(0)+t\mathbf{K}(1)\\
		\mathbf{\tilde{M}}(t) &:=(1-t)\mathbf{M}(0)+t\mathbf{M}(1)\;.
	\end{align}
\end{subequations}
They also allow to match eigenvalues from $t=0$ to $t=1$. 
The construction in \eqref{eq:homotopy} exploits the fact that the discretizations of $\Omega^{(1)}$ and $\Omega^{(2)}$ and thus also the matrix dimensions are compatible. %since they are mapped from the same control mesh in the reference domain. 
The advantage is that the matrix-valued functions $\mathbf{\tilde{K}}(t)$ and $\mathbf{\tilde{M}}(t)$ are linear in $t$, while $\mathbf{K}(t)$ and $\mathbf{M}(t)$ have a nonlinear dependence. 
This will reduce numerical costs for matrix assembly and makes derivative computations, e.g. $\mathrm{d}/\mathrm{d}t\;\mathbf{\tilde{M}}(t)$, trivial. 
However, the original functions \eqref{eq:matrix-def} always correspond to \emph{physical} geometries, while \eqref{eq:homotopy} are purely \emph{algebraic} and have no geometric counterpart for $0<t<1$.
%------------------------------------------------------------------------
\subsection{Basis Functions}
%------------------------------------------------------------------------
The various FE methods differ in the construction of the approximation space $V_h$, e.g. basis functions of different order, element types etc. We refer to \cite{Monk_2003aa,Boffi_2010aa} for the classical FE method. 
However, we will follow~\cite{Buffa_2010aa} and employ IGA using curl-conforming B-splines. This allows us to use the same basis functions for the discretization as for the geometry description, i.e., without an additional meshing step.

The B-spline space of degree $p_i$ along dimension
$i$, with regularity $C^{\alpha_i}$, on the domain $\hat{\Omega}$
shall be denoted by $S^{p_1,p_2,p_3}_{\alpha_1,\alpha_2,\alpha_3}(\hat{\Omega})$. 
Then, following~\cite{Buffa_2010aa}, 
the discrete space $\hat{V}_h$ is chosen as
\begin{equation}
\hat{V}_h := S^{p_1-1,p_2,p_3}_{\alpha_1-1,\alpha_2,\alpha_3}
  \times S^{p_1,p_2-1,p_3}_{\alpha_1,\alpha_2-1,\alpha_3}
  \times S^{p_1,p_2,p_3-1}_{\alpha_1,\alpha_2,\alpha_3-1}.
\end{equation}
Finally, the discrete space $V_h$ in the physical domain is obtained through a curl-conforming 
transformation~\cite{Monk_2003aa} using our domain transformation $G_t$ from \eqref{eq:iga-map-t}.
In \cite{Buffa_2010aa} it is shown that the B-spline basis functions from these spaces form a discrete
de Rham sequence in both, the reference and physical domain, such that they are suitable for the
discretization of $H(\mathrm{curl};\Omega)$. We employ the open source implementation GeoPDEs in Matlab for the discretization \cite{Vazquez_2016aa,Mathworks_2020aa}.

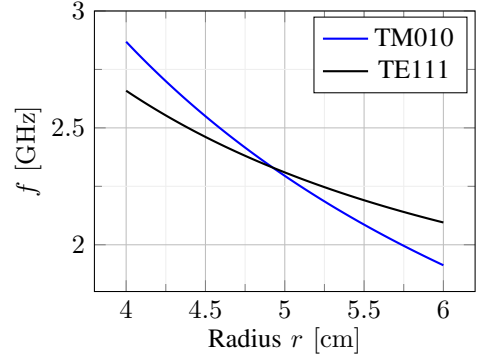
\begin{figure}
	\centering
	% \begin{tikzpicture}
% \begin{axis}[width=0.75\linewidth, height=0.6\linewidth,
% 	xlabel style = {yshift=.5em},
% %	legend pos=south east, 
% %	legend style={font=\footnotesize},
% 	xlabel={Radius $r~ [\mathrm{m}]$}, 
% 	title style = {yshift=-.5em, xshift=4em},
% 	ylabel={$f ~ [\mathrm{Hz}]$ },
% %	ylabel style = {yshift=-1em},
% 	grid=both,
% 	minor grid style={gray!25},
% 	major grid style={gray!25},
% 	clip mode = individual]

% \addplot[color=blue, mark=*, mark size= 1pt] table[x=r,y=f,col sep=comma]{data/Pillbox_frequ_TM010_radii.csv};
% \addplot[color=black, mark=*, mark size= 1pt] table[x=r,y=f,col sep=comma]{data/Pillbox_frequ_TE111_radii.csv};
% %\addplot[only marks, red, mark=*, mark size= 1pt] coordinates {(0.06,2095384106)};
% \node at (axis cs:0.06,2095384106) [anchor=west] {{\scriptsize TE111}};
% %\addplot[only marks, red, mark=*, mark size= 1pt] coordinates {(0.06,1912375463)};
% \node at (axis cs:0.06,1912375463) [anchor=west] {{\scriptsize TM010}};
% \end{axis}
% \end{tikzpicture}

\begin{tikzpicture}
\begin{axis}[
    ymin = 1.8e9, ymax = 3.0e9,
    xtick distance = 0.005,
    % ytick distance = 0.05,
    xlabel style = {yshift=.5em},
    ylabel style = {yshift=-1em},
    grid = both,
    minor tick num = 1,
    major grid style = {lightgray},
    minor grid style = {lightgray!25},
    width=0.75\linewidth, height=0.6\linewidth,
    xlabel = {Radius $r~[\mathrm{cm}]$},
    ylabel = {$f~[\mathrm{GHz}]$},
    scaled y ticks=false,
    yticklabel={$\pgfmathfloatparsenumber{\tick}
        \pgfmathfloattomacro{\pgfmathresult}{\F}{\M}{\E}
        \pgfmathprintnumber{\M}$},
    scaled x ticks=false,
    xticklabel={$\pgfmathfloatparsenumber{\tick}
        \pgfmathfloattomacro{\pgfmathresult}{\F}{\M}{\E}
        \pgfmathprintnumber{\M}$}
    ]
 
% Plot a function
\addplot[
    domain = 0.04:0.06,
    samples = 100,
    smooth,
    thick,
    blue,
] {(2.997924580000066e+08 / (2*pi) * sqrt((2.40482555769577/x)^2 + (0*pi/0.1)^2))}; 
\addplot[
    domain = 0.04:0.06,
    samples = 100,
    smooth,
    thick,
    black,
] {(2.997924580000066e+08 / (2*pi) * sqrt((1.84118378134066/x)^2 + (1*pi/0.1)^2))};
 
\legend{TM010, TE111};
\end{axis}
 
\end{tikzpicture}
	\vspace*{-2em}
	\caption{Crossing of eigenvalues of $\mathrm{TM}0\myspace1\myspace0$ and $\mathrm{TE}1\myspace1\myspace1$ at $r= \SI{4.92} {\centi \meter}$ demonstrated for pillbox cavity.}
	\label{fig:crossing}
\end{figure}
%------------------------------------------------------------------------
\section{Shape Morphing}\label{sec:classification}
%------------------------------------------------------------------------
To classify eigenmodes, we propose to follow the eigenvalues and eigenvectors along $t\in T=[0,1]$ from the complex geometry to a simple geometry, e.g. pillbox. This allows to use the established nomenclature from field theory \cite[Section 8.7]{Jackson_1998aa} and is less error-prone than methods based on counting zero-crossing or maxima \cite{Brackebusch_2014aa} as we will demonstrate in Section~\ref{sec:application}. However, our classification method requires a numerical tracking procedure of the eigenpairs \cite{Lui_1997aa, Jorkowski_2018aa, Safin_2016aa, Georg_2019aa}. This is necessary since eigenvalues may cross on $T$, see Fig.~\ref{fig:crossing}, such that a numbering with respect to the magnitude of the eigenvalues at a given $t$ is not reliable. 

%------------------------------------------------------------------------
\subsection{Eigenvalue Tracking}\label{sec:tracking}
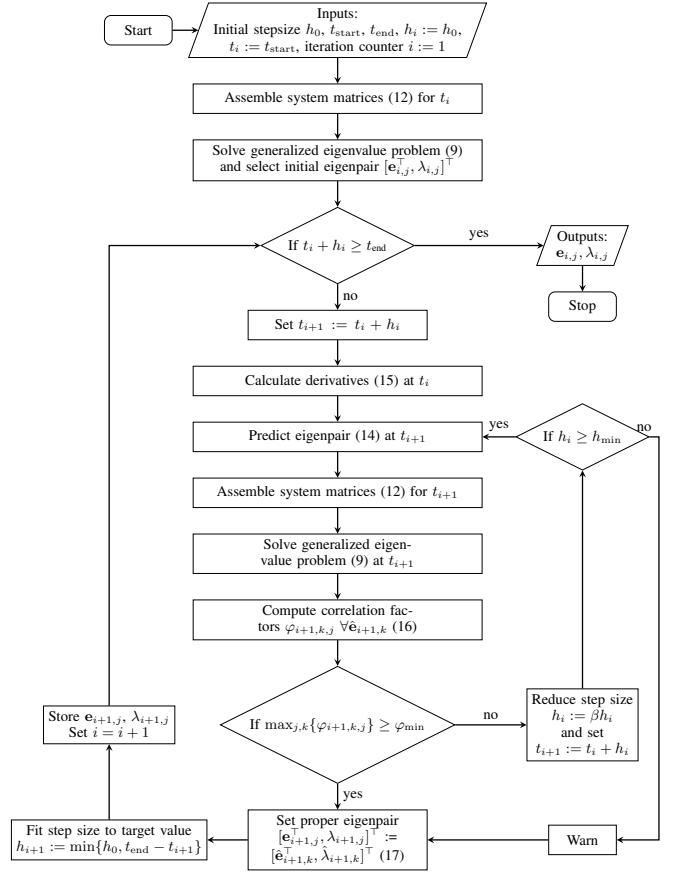
\begin{figure}[t]
	\centering
	\scalebox{0.62}{\begin{tikzpicture}[node distance=1.75em, on grid=false]

\small

\newcommand*{\Breite}{.33\textwidth};
\newcommand*{\BreiteZwei}{.2\textwidth};

\node (start) [startstop] {Start};

\node (in1) [io, right =of start, align=center] {Inputs: \\ Initial stepsize $h_0$,   $t_\mathrm{start}$, $t_\mathrm{end}$,  $h_i := h_0$,\\ $t_i := t_{\mathrm{start}}$, iteration counter $i := 1$};

\node (assemble_init) [process, below = of in1, align=center, text width=\Breite] {Assemble system matrices \eqref{eq:homotopy} for $t_i$};

\node (solve_init) [process, below = of assemble_init, align=center, text width=\Breite] {Solve generalized eigenvalue problem~\eqref{eq:eigprob} \\ and select initial eigenpair $[\mathbf{e}_{i,j} ^\top, \lambda_{i,j} ]^\top$};

%\node (IncreasePinit) [process, below =of  solve_init, text width=\BreiteZwei]{};%Set $h_{i+1} = h_i$};% \\ and $i := i +1$};

\node (dec_finished) [decision, below = of solve_init] {If $t_{i} + h_i\geq t_\textrm{end}$};

\node (IncreaseT) [process, below =of  dec_finished, text width=\BreiteZwei]{Set $t_{i+1} := t_{i} +h_i$};

\node (get_derivatives) [process, align=center, below = of IncreaseT, text width=\Breite] {Calculate derivatives \eqref{eq:eigprobderiv} at $t_{i}$};

\node(Predict) [process, align=center, below =of get_derivatives, text width=\Breite] {Predict eigenpair \eqref{eq:prediction} at $t_{i+1}$ };

\node (assemble) [process, below = of Predict, align=center, text width=\Breite] {Assemble system matrices \eqref{eq:homotopy} for $t_{i+1}$};

\node (solve_gevp) [process, below =of assemble, align=center, text width=\Breite] {Solve generalized eigenvalue problem~\eqref{eq:eigprob} at $t_{i+1}$};

\node (comp_corr) [process, below =of solve_gevp, align=center, text width=\Breite] {Compute correlation factors $\varphi_{i+1,k,j}~\forall \hat{\mathbf{e}}_{i+1,k}$ \eqref{eq:corrfactor}};

\node (dec_corr) [decision, below =of comp_corr] {If $\max_{j,k} \{\varphi_{i+1,k,j}\} \geq \varphi_{\min}$};

% \node (IncreaseP) [process, below =of  dec_corr, text width=\BreiteZwei]{Set $t_{i+1}:=t_i+h_i$ \\ and $i := i +1$};

\node (SetEP) [process, below =of  dec_corr, text width=\BreiteZwei]{Set proper eigenpair $[\mathbf{e}_{i+1,j} ^\top, \lambda_{i+1,j}]^\top$ := $[\hat{\mathbf{e}}_{i+1,k} ^\top, \hat{\lambda}_{i+1,k}]^\top$ \eqref{eq:setEigenpair_i}};

\node (dec_stepsize) [decision, right =of  Predict, xshift=.5em,align=center]  {If $h_i \geq h_{\min}$};

\node (DecreaseStepsize) [process,align = center] at(dec_corr -| dec_stepsize) {Reduce step size \\ $h_i := \beta h_i$ \\ and set \\ $t_{i+1} := t_{i} + h_i$};

\node(Warn) [process, align = center] at (SetEP -| DecreaseStepsize) {Warn};

\node(FitStepsize) [process, left =of SetEP,  xshift=-1em,align=center]{Fit step size to target value\\
	$h_{i+1}:=\min\{h_0, t_\mathrm{end}-t_{i+1}\}$};

\node(Store) [process, align=center] at(FitStepsize |- dec_corr) {Store $\mathbf{e}_{i+1,j}$, $\lambda_{i+1,j}$\\ Set $i = i + 1$};

\node(outputs) [io, align=center] at(DecreaseStepsize |- dec_finished) {Outputs: \\$\mathbf{e}_{i,j}, \lambda_{i,j}$};

\node (stop) [startstop, below = of outputs] {Stop};

\draw[arrow] (start) -- (in1);
\draw[arrow] (in1) -- (assemble_init);
\draw[arrow] (assemble_init) -- (solve_init);
\draw[arrow] (solve_init) --(dec_finished);
\draw[arrow] (dec_finished) --node[anchor = south] {yes} (outputs);
\draw[arrow] (outputs) -- (stop);
\draw[arrow] (dec_finished)  --node[anchor=west] {no}  (IncreaseT);
\draw[arrow] (IncreaseT)   -- (get_derivatives);
\draw[arrow] (get_derivatives) -- (Predict);
\draw[arrow] (Predict)   -- (assemble);
\draw[arrow] (assemble) -- (solve_gevp);
\draw[arrow] (solve_gevp) -- (comp_corr);
\draw[arrow] (comp_corr) -- (dec_corr);
\draw[arrow] (dec_corr) --node[anchor=south]{no} (DecreaseStepsize);
\draw[arrow] (dec_corr) --node[anchor=west]{yes} (SetEP);
\draw[arrow] (SetEP) -- (FitStepsize);
\draw[arrow] (DecreaseStepsize) -- (dec_stepsize);
\draw[arrow] (dec_stepsize) --node[anchor=south]{yes} (Predict);
\node (ctrl1) [right=of DecreaseStepsize, anchor=north] {};
\draw[arrow] (dec_stepsize) -| node[anchor=south, xshift=-1em]{no}(ctrl1.west)  |- (Warn);
\draw[arrow] (Warn) -- (SetEP);
\draw[arrow] (FitStepsize) -- (Store);
\draw[arrow] (Store) |- (dec_finished);
\end{tikzpicture}}
	\caption{Flowchart describing the proposed eigenvalue tracking method. % 
	The parameter of the stepsize control is chosen as~$\beta = 0.5$.}
\label{fig:Flowchart_Tracking}
\end{figure}

%------------------------------------------------------------------------
We discretize the parameter interval $T$ into $0=t_1<t_2<\ldots<t_i<\ldots<t_M=1$. 
Then we compute sequentially for each $i$ the relevant eigenpairs  $j\in J\subset\{1,\ldots,N\}$ by solving \eqref{eq:eigprob} on geometry $\Omega_{t_i}$. 
Let us assume we have solved the eigenvalue problem \eqref{eq:eigprob} at step $i$
by an eigenvalue solver, e.g. an Arnoldi or Lanczos method~\cite{Lehoucq_1998aa}. 
The solutions are denoted by $\mathbf{e}_{i,j} := \mathbf{e}_j\bigl(t_i\bigr)$ and $\lambda_{i,j} := \lambda_j(t_i)$. 
We now aim at computing the eigenpairs for step $i+1$ but
since an eigenvalue crossing may occur between $t_i$ and $t_{i+1}$, the consistent sorting of the eigenpairs at $t_{i+1}$ must be determined. 
We denote the unsorted candidates from solving the eigenvalue problem~\eqref{eq:eigprob} on $\Omega_{t_{i+1}}$ by $\hat{\mathbf{e}}_{i+1,k}$ and $\hat{\lambda}_{i+1,k}$. They are compared with first-order extrapolations from the current step, i.e.,
\begin{align}
	\begin{bmatrix}
		\mathbf{\tilde{e}}_{i+1,j}\\
		\tilde{\lambda}_{i+1,j}
	\end{bmatrix}
	=
	\begin{bmatrix}
		\mathbf{e}_{i+1}\\
		\lambda_{i+1}
	\end{bmatrix}
	+h_i
	\begin{bmatrix}
		\mathbf{e}_{i,j}'\\
		\lambda_{i,j}'
	\end{bmatrix} 
	\label{eq:prediction}
\end{align}
to establish the matching where we used $h_i:=t_{i+1}-t_{i}$ for the stepsize. In the case of non-degenerated eigenpairs the derivatives are easily obtained by solving
\begin{align}
	\label{eq:eigprobderiv}
	\begin{bmatrix}
		\mathbf{K}_{i} - \lambda_{i,j} \mathbf{M}_{i}  & -\mathbf{M}_{i}\mathbf{e}_{i,j}
		\\
		\mathbf{e}_{i-1,j}^{\mathsf{H}}\mathbf{M}_{i} & 0
	\end{bmatrix}
	&\begin{bmatrix}
		\mathbf{e}_{i,j}'
		\\
		\lambda_{i,j}'
	\end{bmatrix}
	\\
	=&
	\begin{bmatrix}
		-\mathbf{K}_{i}'\mathbf{e}_{i,j}
		+
		\lambda_{i,j}\mathbf{M}_{i}'\mathbf{e}_{i,j}
		\\
		-\mathbf{e}_{i,j}^{\mathsf{H}} \mathbf{M}_{i}'\mathbf{e}_{i,j}
	\end{bmatrix}
	\nonumber
\end{align}
where we used the definitions $\mathbf{K}_{i} := \mathbf{K}(t_i)$, $\mathbf{M}_i := \mathbf{M}(t_i)$ and added a normalization constraint based on $\mathbf{e}_{i-1,j}$.
Finally, we use the correlation factor proposed by \cite{Jorkowski_2018aa}
\begin{equation}
	\varphi_{i+1,k,j} 
	= 
	\frac{
		\hat{\mathbf{e}}_{i+1,k}^{\mathsf{H}}
		\mathbf{M}_{i+1}
		\tilde{\mathbf{e}}_{i+1,j}
	}{
		\|\hat{\mathbf{e}}_{i+1,k}\|_{\mathbf{M}_{i+1}} 		\|\tilde{\mathbf{e}}_{i+1,j}\|_{\mathbf{M}_{i+1}}
	}
	\label{eq:corrfactor}
\end{equation}
with the norm $\|\mathbf{x}\|_\mathbf{A} = \sqrt{\mathbf{x}^{\mathsf{H}} \mathbf{A} \mathbf{x}}$ for the matching. 
The correlation factor compares the normalized scalar products of the predicted eigenvectors ${\tilde{\mathbf{e}}}_{i+1,j}$ and the new candidates $\hat{\mathbf{e}}_{i+1,k}$ that are obtained by solving the eigenvalue problem at $t_{i+1}$. In the ideal case, i.e., the eigenpairs depend linearly on $t$, we find $j$ and $k$ such that $\hat{\mathbf{e}}_{i+1,k} = \tilde{\mathbf{e}}_{i+1,j}$. Then, the correlation factor is  $\varphi_{i+1,k,j} = 1$ and $\varphi_{i+1,l,j} = 0$ for all other $l\neq k$ since it is a product of orthonormal vectors. In the general case, all $\varphi_{i+1,k,j}$ will attain values between $0$ and $1$ and a correlation is computed for all pairs $j,k$. We match the one with the highest correlation factor and define
\begin{align}
	\begin{bmatrix}
		\mathbf{e}_{i+1,j}\\
		\lambda_{i+1,j}
	\end{bmatrix}
	:=
	\begin{bmatrix}
		\hat{\mathbf{e}}_{i+1,k}\\
		\hat{\lambda}_{i+1,k}
	\end{bmatrix}
	\text{ with }
	\arg \max_{j,k} \left\{
		\varphi_{i+1,k,j}
	\right\}\;.
	\label{eq:setEigenpair_i}
\end{align}
To reduce the risk that an eigenpair is wrongly matched, e.g. due to an insufficiently small step size, the new eigenpair is only accepted if it fulfills the criterion
$$\max_{j,k} \varphi_{i+1,k,j} \geq \varphi_{\min}$$
with respect to a user-specified threshold correlation factor~$\varphi_{\min}$. 
If the criterion is not met, the step is repeated with a smaller stepsize. The tracking algorithm is illustrated by the flowchart in Fig.~\ref{fig:Flowchart_Tracking}.

Note that the presented version of the algorithm requires non-degenerated eigenmodes at each $t_i$. However, the algorithm can be generalized by using Ojalvo's method for computing the derivatives instead of \eqref{eq:eigprobderiv}, \cite{Ojalvo_1986aa, Dailey_1989aa}. 
However, this was not necessary in our computations.

%------------------------------------------------------------------------
\subsection{Identification of Eigenmodes}
%------------------------------------------------------------------------
We use the eigenvalue tracking as proposed in section~\ref{sec:tracking} to recognize eigenmodes by tracking the eigenvalues from the complex geometry $\Omega^{(1)}$ without known classification to a geometry $\Omega^{(2)}$ where the eigenmodes, eigenfrequencies and classifications are known, e.g. the pillbox cavity. After having solved the eigenvalue problem~\eqref{eq:eigprob} at~$t=0$, we choose an eigenpair~$j$ for which we want to convey the classification system. 
We apply the presented algorithm and track the eigenmodes to the geometry $\Omega^{(2)}$ and finally match the result with known closed-form eigenfrequencies, see e.g. (\ref{eq:resonanceFrequenciesTM}-\ref{eq:resonanceFrequenciesTE}). We can use either the physical \eqref{eq:matrix-def} or algebraic mapping \eqref{eq:homotopy}. Both yield the same results as will be demonstrated later in Section~\ref{sec:application}.

\subsection{Exploration of Eigenmodes}\label{sec:backwardTracking}
% "Backward Tracking": Pillbox->TESLA
The same tracking algorithm can also be used `backwards', i.e. from $t=1$ down to $t=0$, to explore a specific eigenmode $m,n,p$ in the complex geometry $\Omega^{(1)}$ for which the number~$k$ of the eigenmode~$\mathbf{e}_k$ is not known.
To specify the eigenmode that should be detected, the known eigenfrequency and corresponding classification are chosen and matched with the numerical solution of the eigenvalue problem~\eqref{eq:eigprob} on $\Omega^{(2)}$.
The eigenmode~$\mathbf{e}_{1,j}$ with $m,n,p$ on $\Omega^{(2)}$ is then tracked to the geometry $\Omega^{(1)}$, in which it is to be analyzed.

This procedure allows to compute a single eigenmode in an efficient way. Otherwise possibly many eigenpairs in the assumed vicinity of the desired eigenmode for the geometry $\Omega^{(1)}$ have to computed and then forwardly tracked until the desired one is identified on $\Omega^{(2)}$.

%------------------------------------------------------------------------
\section{Application}\label{sec:application}
%------------------------------------------------------------------------
In this section, we present the application of the identification and exploration to the TESLA cavity and the cylindrical pillbox cavity. The classifications of the eigenmodes of the pillbox cavity into $\mathrm{TE}$ and $\mathrm{TM}$ modes and according to the field distribution in azimuthal, radial and longitudinal direction are known analytically and are given in \eqref{eq:resonanceFrequenciesTM} and \eqref{eq:resonanceFrequenciesTE}.

Following the theory of electromagnetic fields, the azimuthal index~$m$ can be related to the number of zero crossings resp. of local field extrema in azimuthal direction and  indicates the number of poles of the mode according to a classical multipole expansion, where $m = 0$ refers to a monopole mode, $m=1$ to a dipole mode, $m=2$ to a quadrupole mode and so forth.
The radial index~$n$ corresponds to the number of half-waves in radial direction and the longitudinal index~$p$ analogously in longitudinal direction.

The goal is to convey this nomenclature to the TESLA cavity without relying on error-prone counting of zero-crossings or maxima and minima.
In order to perform this eigenmode classification automatically, the eigenmodes of the TESLA cavity are tracked along the deformation to the pillbox cavity and matched with the analytical solutions. 
%------------------------------------------------------------------------
\subsection{TESLA Cavity}
%------------------------------------------------------------------------
The TESLA cavity is a radio-frequency superconducting cavity designed for linear accelerators. We use here the design that is installed at DESY in Hamburg, \cite{Edwards_1995aa}. It is a 9-cell standing wave structure of about \SI{1}{\metre} in length, whose accelerating eigenmode resonates at \SI{1.3}{\giga\hertz}. 
Following~\cite{Corno_2017ad}, instead of the \SI{141.6}{\mm} beampipes with connected higher-order mode (HOM) couplers, cylindrical beampipes with an increased length of \SI{365}{\mm} are attached to the model, such that the effects of the boundary conditions at the end caps of the beampipes on the eigenmodes can be neglected. 
This approximation is valid for eigenmodes below the cutoff frequency of approximately \SI{2.2}{\giga \hertz}, this includes the first monopole and the first two dipole passbands.

In order to achieve the desired field flatness within the cavity, we conduct a tuning procedure before tracking the eigenmodes, following~\cite{Corno_2017ad}.
We employ two criteria to quantify the field flatness and combine them to a weighted objective function.
We denote the peak value of the longitudinal component of the electric field at radius $r=0$ by $E_{\mathrm{peak},q}$ in the $q$th cell.
To improve the field flatness, we modify the length of the two end-cups of the TESLA cavity and the radius of the cell-equator using a particle swarm optimizer until we reach field flatness with
\begin{equation}
    \eta_1 = 1-\frac{(\max_q |E_{\mathrm{peak},q}| - \min_q |E_{\mathrm{peak},q}|)}{\mathbb{E}(|E_{\mathrm{peak},q}|)}
\end{equation}
and 
\begin{equation}
    \eta_2 = 1- \frac{\mathrm{std} (E_{\mathrm{peak},q})}{\mathbb{E}(|E_{\mathrm{peak},q}|)}
\end{equation}
of $\eta_1,\eta_2\geq 0.95$. 
 The parameters for the construction of the TESLA cavity and our tuning values are shown in Tab.~\ref{tab:tesla_tuning}.
 With those tuning values, we achieve a field flatness with $\eta_1 = 0.9907$ and $\eta_2 = 0.9965$.
 
\begin{table}
    \centering
    \caption{Parameters of the TESLA half-cells and tuning, all dimensions in $\mathrm{mm}$~\cite{Aune_2000aa} and the difference due to tuning, cf. Fig.~\ref{fig:tesla_shape}}
    \label{tab:tesla_tuning}
    \begin{tabular}{l|lll}
        cavity shape parameter & midcup  & endcup 1 &  endcup 2 \\ 
         & (+tuning) & (+tuning) & (+tuning) \\ \hline
        equator radius $R_\mathrm{{eq}}$ & 103.3 (+ 0.89) & 103.3 & 103.3 \\
        iris radius $R_\mathrm{{iris}}$ & 35 & 39 & 39 \\
        radius $R_\mathrm{{arc}}$ of arc & 42.0 & 40.3 & 42 \\
        horizontal half axis $a$ & 12 & 10 & 9 \\
        vertical half axis $b$ & 19 & 13.5 & 12.8 \\
        length $l$ & 57.7 & 56.0 (+0.90) & 57.0 (+1.04) \\ 
    \end{tabular}
 \end{table}

Solving~\eqref{eq:eigprobderiv} requires the derivatives $\mathbf{K}_i'$ and~$\mathbf{M}_i'$.
While the computation of the derivatives for the algebraic matrix mappings~\eqref{eq:homotopy} is trivial due to their linear construction, we compute the derivatives for the geometric matrices~\eqref{eq:matrix-def} by finite differences
\begin{subequations}
\begin{align}
\mathbf{K}'(t)&:=\frac{\mathrm{d}}{\mathrm{d}t} \mathbf{K}(t) \approx \frac{1}{\delta}\left(\mathbf{K}(t + \delta)-\mathbf{K}(t)\right) \\
\mathbf{M}'(t)&:=\frac{\mathrm{d}}{\mathrm{d}t} \mathbf{M}(t) \approx \frac{1}{\delta}\left( \mathbf{M}(t+\delta)-\mathbf{M}(t)\right),
\end{align}
\end{subequations}
with sufficiently small $\delta$.

We start our algorithm with an initial stepsize of $h_0 = 0.1$ and decrease the stepsize by $\beta = 0.5$ if the maximum correlation factor is smaller than $\varphi_{\min} = 0.9$.
If the stepsize $h_i$ gets smaller than e.g. $h_{\min} = 0.00125$, we accept the current eigenpair nevertheless and issue a warning.
For the computation of the finite differences, we use $\delta=10^{-6}$.

%--------------------------------------------------------------
\subsection{Shape morphing}
%--------------------------------------------------------------
In order to classify the eigenmodes of the TESLA cavity by applying the eigenvalue tracking as described in section~\ref{sec:tracking}, we consider a linear shape morphing from the \mbox{9-cell} TESLA cavity to a pillbox cavity of radius $r=\SI{3.9}{\centi\meter}$ and length $l=\SI{103.61}{\centi\meter}~(+\SI{0.19}{\centi \meter})$.
This corresponds to the iris radius of the TESLA cavity endcups and the length of the (tuned) TESLA cavity without the beampipes, respectively, see Tab.~\ref{tab:tesla_tuning}.
The CAD data are given for download in \cite{Ziegler_2022aa}.
The model of the cavity is discretized using second degree basis functions and \num{24960} degrees of freedom.
The morphing of the shape of the \mbox{9-cell} TESLA cavity (using the physical mapping) and the concomitant deformation of the magnitude of the electric field within the cavity is illustrated in Fig.~\ref{fig:deformationTM018} from the original TESLA cavity at deformation parameter $t = 0$ in Fig.~\ref{fig:0000} to the pillbox cavity at $t=1$ in Fig.~\ref{fig:0020} for the accelerating eigenmode $\mathrm{TM}0\myspace1\myspace8$.

%------------------------------------------------------------------------
\subsection{Identification of Eigenmodes}
%------------------------------------------------------------------------
\begin{figure}
	\centering
	\input{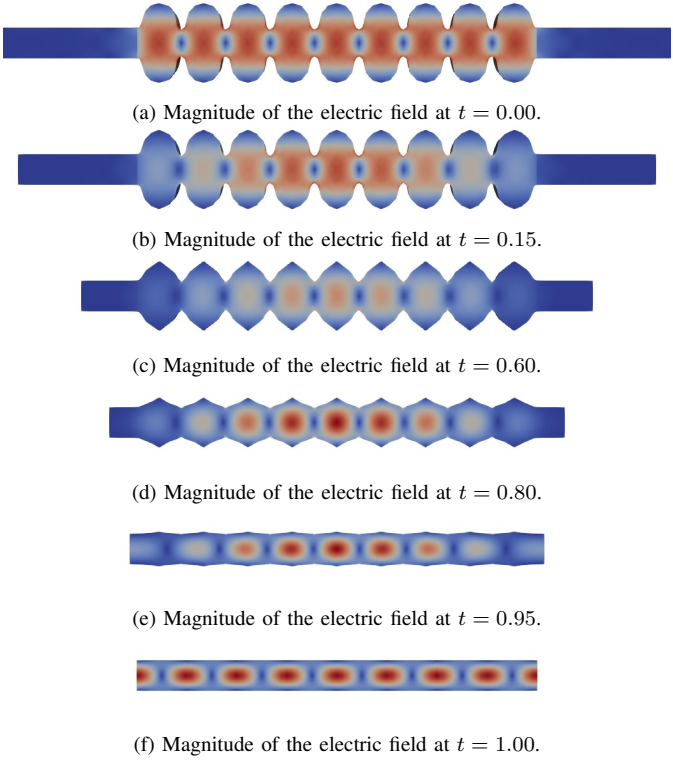}
	\caption{Electric field magnitude of the accelerating eigenmode ($\mathrm{TM}0\myspace1\myspace8$) along the physical deformation.}
	\label{fig:deformationTM018}
\end{figure}

The eigenmode tracking algorithm is applied to the first 45 eigenmodes of the TESLA cavity, corresponding to the first monopole passband and the first and second dipole passband.
The results are presented in Table~\ref{tab:results45modes}, ordered by the eigenfrequency.

\begin{table}
	\caption{Identification results for the fist 45 eigenmodes of the TESLA cavity. The different passbands are separated by horizontal lines.}
	\footnotesize	
			\centering
		\pgfplotstabletypeset[
  col sep=comma,
  % change ordering of cols:
  columns={type, m, n, p},
  columns/type/.style={string type,column type=c},
  columns/m/.style={string type,column type=c,column name={\makecell[t]{$m$ \\ $\phantom{x}$}} },
  columns/n/.style={string type,column type=c,column name=$n$},
  columns/p/.style={string type,column type=c,column name=$p$},
  every head row/.style={before row=\toprule,after row=\midrule},
			every last row/.style={after row=\bottomrule},
			every row no 9/.style={before row=\toprule},
			every row no 27/.style={before row=\toprule},%
  ]
{	type, m, n, p
	$\mathrm{TM}$,$0$,$1$,$0$
	$\mathrm{TM}$,$0$,$1$,$1$
	$\mathrm{TM}$,$0$,$1$,$2$
	$\mathrm{TM}$,$0$,$1$,$3$
	$\mathrm{TM}$,$0$,$1$,$4$
	$\mathrm{TM}$,$0$,$1$,$5$
	$\mathrm{TM}$,$0$,$1$,$6$
	$\mathrm{TM}$,$0$,$1$,$7$
	$\mathrm{TM}$,$0$,$1$,$8$
	$\mathrm{TE}$,$1$,$1$,$1$
	$\phantom{x}$, , ,
	$\mathrm{TE}$,$1$,$1$,$2$
	$\phantom{x}$, , ,
	$\mathrm{TE}$,$1$,$1$,$3$
	$\phantom{x}$, , ,
	$\mathrm{TE}$,$1$,$1$,$4$
	$\phantom{x}$, , ,
	$\mathrm{TE}$,$1$,$1$,$5$
	$\phantom{x}$, , ,
	$\mathrm{TE}$,$1$,$1$,$6$
	$\phantom{x}$, , ,
	$\mathrm{TE}$,$1$,$1$,$7$
	$\phantom{x}$, , ,
	$\mathrm{TE}$,$1$,$1$,$8$
	$\phantom{x}$, , ,
	$\mathrm{TE}$,$1$,$1$,$9$
	$\phantom{x}$, , ,
	$\mathrm{TE}$,$1$,$1$,$10$
	$\phantom{x}$, , ,
	$\mathrm{TE}$,$1$,$1$,$11$
	$\phantom{x}$, , ,
	$\mathrm{TE}$,$1$,$1$,$12$
	$\phantom{x}$, , ,
	$\mathrm{TE}$,$1$,$1$,$13$
	$\phantom{x}$, , ,
	$\mathrm{TE}$,$1$,$1$,$14$
	$\phantom{x}$, , ,
	$\mathrm{TE}$,$1$,$1$,$15$
	$\phantom{x}$, , ,
	$\mathrm{TE}$,$1$,$1$,$16$
	$\phantom{x}$, , ,
	$\mathrm{TE}$,$1$,$1$,$17$
	$\phantom{x}$, , ,
	$\mathrm{TE}$,$1$,$1$,$18$
	$\phantom{x}$, , ,
}~\begin{filecontents*}{data/freq_table_tuning.csv}
1,1.2763,1.2758786719367,3.30116793310755
2,1.2784,1.2779720520217,3.34752799043145
3,1.2816,1.2811941463504,3.16677317101969
4,1.2855,1.28517167909275,2.55403272848378
5,1.2898,1.28943557207591,2.82546072325275
6,1.2938,1.29347234928654,2.53246802794987
7,1.2971,1.29678527069598,2.42640740128658
8,1.2993,1.29895869542509,2.62683425625101
9,1.3,1.29972813454941,2.09127269684535
10,1.6205,1.62349213299969,18.4642579431631
11,1.6206,1.62349213299969,17.8460631845734
12,1.628,1.63057625668032,15.8246724835221
13,1.6282,1.63057625668032,14.5943783338511
14,1.6408,1.6428611824938,12.5620581045565
15,1.641,1.6428611824938,11.3417580365641
16,1.6585,1.66025797645088,10.5997977140894
17,1.6587,1.66025797645089,9.39275607938411
18,1.6806,1.6821939615602,9.48447911578474
19,1.6809,1.6821939615602,7.69802820036161
20,1.7061,1.7076994099291,9.37465523180848
21,1.7063,1.7076994099291,8.20142957922625
22,1.7335,1.73547919426965,11.4173306584805
23,1.7338,1.73547919426966,9.68505173412515
24,1.7614,1.76401698969066,14.8574411868954
25,1.7616,1.76401698969067,13.7204228580066
26,1.7886,1.79225586284663,20.4398012223626
27,1.789,1.79225586284664,18.1993451461198
28,1.7992,1.80635818845533,39.7853960389651
29,1.7995,1.80635818845533,38.1116335389309
30,1.8375,1.83959825682699,11.4190847727289
31,1.8375,1.83959825682701,11.4190847728548
32,1.8531,1.8569693077706,20.8801887140266
33,1.8532,1.85696930777064,20.3394548383434
34,1.8655,1.87055264754553,27.0846826348577
35,1.8655,1.87055264754555,27.0846826349344
36,1.8746,1.88056591858184,31.8250217744727
37,1.8746,1.88056591858191,31.8250217748212
38,1.8809,1.88758576424823,35.5455592973056
39,1.8809,1.88758576424823,35.5455592973056
40,1.885,1.89223080560724,38.3597114442413
41,1.885,1.89223080560725,38.3597114443121
42,1.8875,1.89505240757508,40.0127553646833
43,1.8875,1.89505240757514,40.0127553649612
44,1.8887,1.89651080839803,41.3554741252199
45,1.8887,1.89651080839803,41.3554741252199
\end{filecontents*}
\pgfplotstabletypeset[
			columns={[index]0,[index]1,[index]2,[index]3},
			col sep=comma,
			columns/0/.style={
				column name=Mode
			},
			columns/1/.style={
				column name={\makecell[t]{Freq.\\ (GHz) \cite{Ackermann_2015aa}}},
				precision=4
			},
			columns/2/.style={
				column name={\makecell[t]{Freq.\\ (GHz) }},
				precision=4
			},
			columns/3/.style={
				column name={\makecell[t]{Rel. Error \\ ($\cdot10^{-4}$)}},
				precision=6
			},
			every head row/.style={before row=\toprule,after row=\midrule},
			every last row/.style={after row=\bottomrule},
			every row no 9/.style={before row=\toprule},
			every row no 27/.style={before row=\toprule},%
		]{data/freq_table_tuning.csv}
	\label{tab:results45modes}
\end{table}

As one can see, the first nine eigenmodes are monopole modes with radial index $n = 1$ and longitudinal index $p$ from $0$ to $8$.
The accelerating eigenmode is the $9$th eigenmode, which is classified as $\mathrm{TM}0\myspace1\myspace8$, for which the magnitude of the electric field can be seen in Fig.~\ref{fig:deformationTM018}.
These and further visualizations are accessible in~\cite{Ziegler_2022aa}.
The next 18 eigenmodes correspond to the first dipole mode passband, where each eigenmode appears twice, with different polarisation. 
Eigenmodes $j=28,\ldots,45$ belong to the second dipole passband with longitudinal index $p$ from $10$ to $18$.
We note that our classification of the second dipole passband differs from the established nomenclature, as e.g. in~\cite{Wanzenberg_2001aa}. 
However, this is not surprising since the longitudinal components of the corresponding modes in the TESLA cavity do not allow an unambiguous assignment of TE or TM since both $\mathbf{E}_z$ and $\mathbf{H}_z$ components are not identically zero.

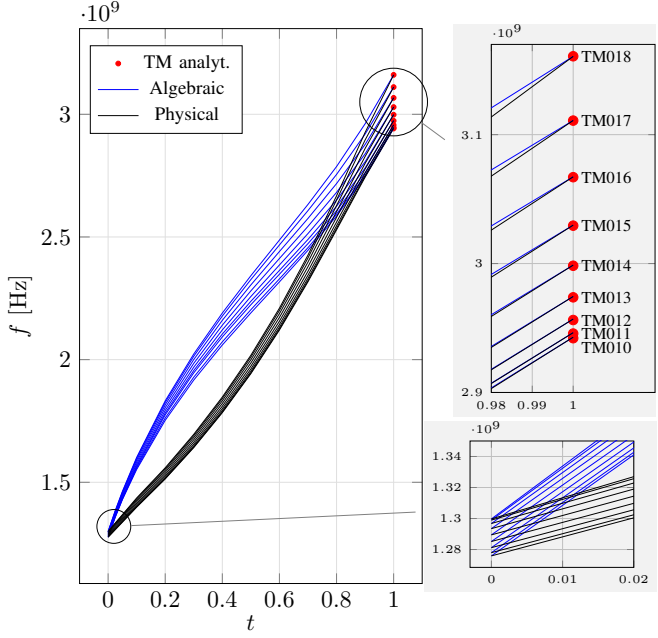
\begin{figure}
	\centering
	\scalebox{0.9}{
\begin{tikzpicture}
\begin{axis}[width=0.75\linewidth, height=.4\textheight,
	xlabel={t},ylabel={Frequency [Hz]}, 
	xlabel style = {yshift=.5em},
	legend pos=north west, 
	legend style={font=\footnotesize},
	xlabel={$t$}, 
	xmax=1.1,
	xmin=-0.1,
	xtick = {0, 0.2, 0.4, 0.6, 0.8, 1},
	title style = {yshift=-.5em, xshift=4em},
	ylabel={$f ~ [\mathrm{Hz}]$ },
	ylabel style = {yshift=-1em},
	grid=both,
	minor grid style={gray!25},
	major grid style={gray!25},
	no marks, clip mode = individual]

\addplot[only marks, red, mark=x, mark size= 1pt] coordinates {				(1,2.9421e9) 
	(1,2.9457e9)
	(1,2.9562e9)
	(1,2.9738e9)
	(1,2.9983e9)
	(1,3.0294e9)
	(1,3.0670e9)
	(1,3.1109e9)
	(1,3.1608e9)};

\addplot[color=blue] %
	table[x=t,y=f,col sep=comma]%
{./data/9cells_modes10_45/Morphing_Mode1_algeb.csv};
\addplot[color=black] table[x=t,y=f,col sep=comma]{./data/9cells_modes10_45/Morphing_Mode1_new_physic.csv};

\addplot[color=blue] table[x=t,y=f,col sep=comma]{./data/9cells_modes10_45/Morphing_Mode2_algeb.csv};
\addplot[color=black] table[x=t,y=f,col sep=comma]{./data/9cells_modes10_45/Morphing_Mode2_new_physic.csv};

\addplot[color=blue] table[x=t,y=f,col sep=comma]{./data/9cells_modes10_45/Morphing_Mode3_algeb.csv};
\addplot[color=black] table[x=t,y=f,col sep=comma]{./data/9cells_modes10_45/Morphing_Mode3_new_physic.csv};

\addplot[color=blue] table[x=t,y=f,col sep=comma]{./data/9cells_modes10_45/Morphing_Mode4_algeb.csv};
\addplot[color=black] table[x=t,y=f,col sep=comma]{./data/9cells_modes10_45/Morphing_Mode4_new_physic.csv};

\addplot[color=blue] table[x=t,y=f,col sep=comma]{./data/9cells_modes10_45/Morphing_Mode5_algeb.csv};
\addplot[color=black] table[x=t,y=f,col sep=comma]{./data/9cells_modes10_45/Morphing_Mode5_new_physic.csv};

\addplot[color=blue] table[x=t,y=f,col sep=comma]{./data/9cells_modes10_45/Morphing_Mode6_algeb.csv};
\addplot[color=black] table[x=t,y=f,col sep=comma]{./data/9cells_modes10_45/Morphing_Mode6_new_physic.csv};

\addplot[color=blue] table[x=t,y=f,col sep=comma]{./data/9cells_modes10_45/Morphing_Mode7_algeb.csv};
\addplot[color=black] table[x=t,y=f,col sep=comma]{./data/9cells_modes10_45/Morphing_Mode7_new_physic.csv};

\addplot[color=blue] table[x=t,y=f,col sep=comma]{./data/9cells_modes10_45/Morphing_Mode8_algeb.csv};
\addplot[color=black] table[x=t,y=f,col sep=comma]{./data/9cells_modes10_45/Morphing_Mode8_new_physic.csv};

\addplot[color=blue] table[x=t,y=f,col sep=comma]{./data/9cells_modes10_45/Morphing_Mode9_algeb.csv};
\addplot[color=black] table[x=t,y=f,col sep=comma]{./data/9cells_modes10_45/Morphing_Mode9_new_physic.csv};

% {\legend{Algebraic, Physical}
\legend{TM analyt., Algebraic,Physical}

\coordinate (spypoint1) at (axis cs:0.02,1.32e9);
\coordinate (spypoint2) at (axis cs:1,3.05e9);
\end{axis}

\node[pin={[pin distance=4.2cm, yshift = 0.3cm]0:{%
		\begin{tikzpicture}[tight background,background rectangle/.style={fill=gray!10}, show background rectangle, framed]
		\begin{axis}[
		no markers,
		grid=major,
		tiny,
		xmin=-0.003,xmax=0.02,
		ymax=1.35e9, xtick = {0, 0.01, 0.02}, scaled x ticks =false	, xticklabel style={
			/pgf/number format/fixed,
			/pgf/number format/precision=2
		},
		]
	%	\addplot coordinates {(0, 0.0) (0, 0.9) (4, 0.9) (5, 1) (6, 0.9) (80, 0)};
		\addplot[color=blue] %
		table[x=t,y=f,col sep=comma]%
		{./data/9cells_modes10_45/Morphing_Mode1_algeb.csv};
		\addplot[color=black] table[x=t,y=f,col sep=comma]{./data/9cells_modes10_45/Morphing_Mode1_new_physic.csv};
		\addplot[color=blue] table[x=t,y=f,col sep=comma]{./data/9cells_modes10_45/Morphing_Mode2_algeb.csv};
		\addplot[color=black] table[x=t,y=f,col sep=comma]{./data/9cells_modes10_45/Morphing_Mode2_new_physic.csv};
		
		\addplot[color=blue] table[x=t,y=f,col sep=comma]{./data/9cells_modes10_45/Morphing_Mode3_algeb.csv};
		\addplot[color=black] table[x=t,y=f,col sep=comma]{./data/9cells_modes10_45/Morphing_Mode3_new_physic.csv};
		
		\addplot[color=blue] table[x=t,y=f,col sep=comma]{./data/9cells_modes10_45/Morphing_Mode4_algeb.csv};
		\addplot[color=black] table[x=t,y=f,col sep=comma]{./data/9cells_modes10_45/Morphing_Mode4_new_physic.csv};
		
		\addplot[color=blue] table[x=t,y=f,col sep=comma]{./data/9cells_modes10_45/Morphing_Mode5_algeb.csv};
		\addplot[color=black] table[x=t,y=f,col sep=comma]{./data/9cells_modes10_45/Morphing_Mode5_new_physic.csv};
		
		\addplot[color=blue] table[x=t,y=f,col sep=comma]{./data/9cells_modes10_45/Morphing_Mode6_algeb.csv};
		\addplot[color=black] table[x=t,y=f,col sep=comma]{./data/9cells_modes10_45/Morphing_Mode6_new_physic.csv};
		
		\addplot[color=blue] table[x=t,y=f,col sep=comma]{./data/9cells_modes10_45/Morphing_Mode7_algeb.csv};
		\addplot[color=black] table[x=t,y=f,col sep=comma]{./data/9cells_modes10_45/Morphing_Mode7_new_physic.csv};
		
		\addplot[color=blue] table[x=t,y=f,col sep=comma]{./data/9cells_modes10_45/Morphing_Mode8_algeb.csv};
		\addplot[color=black] table[x=t,y=f,col sep=comma]{./data/9cells_modes10_45/Morphing_Mode8_new_physic.csv};
		
		\addplot[color=blue] table[x=t,y=f,col sep=comma]{./data/9cells_modes10_45/Morphing_Mode9_algeb.csv};
		\addplot[color=black] table[x=t,y=f,col sep=comma]{./data/9cells_modes10_45/Morphing_Mode9_new_physic.csv};
		\end{axis}
		\end{tikzpicture}%
}},draw,circle,minimum size=0.5cm] at (spypoint1) {};

\node[pin={[pin distance=0.25cm, yshift = -1.75cm]0:{%
		\begin{tikzpicture}[tight background,background rectangle/.style={fill=gray!10}, show background rectangle, framed]
		\begin{axis}[
		no markers,
		grid=major,
		tiny, clip mode = individual,
		xmin=0.98, xmax=1.02, height=.275\textheight, xtick = {0.98, 0.99, 1.0}%, 1.01, 1.02}
		, ytick = {2.9e9, 3.0e9, 3.1e9}, ymin = 2.9e9, ymax=3.17e9
		]
		\addplot[only marks, red, mark=x, mark size= 2pt] coordinates {(1,2.9421e9) 
	(1,2.9457e9)
	(1,2.9562e9)
	(1,2.9738e9)
	(1,2.9983e9)
	(1,3.0294e9)
	(1,3.0670e9)
	(1,3.1109e9)
	(1,3.1608e9)};
		\addplot[color=blue] %
		table[x=t,y=f,col sep=comma]%
		{./data/9cells_modes10_45/Morphing_Mode1_algeb.csv};
		\addplot[color=black] table[x=t,y=f,col sep=comma]{./data/9cells_modes10_45/Morphing_Mode1_new_physic.csv};
		\addplot[color=blue] table[x=t,y=f,col sep=comma]{./data/9cells_modes10_45/Morphing_Mode2_algeb.csv};
		\addplot[color=black] table[x=t,y=f,col sep=comma]{./data/9cells_modes10_45/Morphing_Mode2_new_physic.csv};
		
		\addplot[color=blue] table[x=t,y=f,col sep=comma]{./data/9cells_modes10_45/Morphing_Mode3_algeb.csv};
		\addplot[color=black] table[x=t,y=f,col sep=comma]{./data/9cells_modes10_45/Morphing_Mode3_new_physic.csv};
		
		\addplot[color=blue] table[x=t,y=f,col sep=comma]{./data/9cells_modes10_45/Morphing_Mode4_algeb.csv};
		\addplot[color=black] table[x=t,y=f,col sep=comma]{./data/9cells_modes10_45/Morphing_Mode4_new_physic.csv};
		
		\addplot[color=blue] table[x=t,y=f,col sep=comma]{./data/9cells_modes10_45/Morphing_Mode5_algeb.csv};
		\addplot[color=black] table[x=t,y=f,col sep=comma]{./data/9cells_modes10_45/Morphing_Mode5_new_physic.csv};
		
		\addplot[color=blue] table[x=t,y=f,col sep=comma]{./data/9cells_modes10_45/Morphing_Mode6_algeb.csv};
		\addplot[color=black] table[x=t,y=f,col sep=comma]{./data/9cells_modes10_45/Morphing_Mode6_new_physic.csv};
		
		\addplot[color=blue] table[x=t,y=f,col sep=comma]{./data/9cells_modes10_45/Morphing_Mode7_algeb.csv};
		\addplot[color=black] table[x=t,y=f,col sep=comma]{./data/9cells_modes10_45/Morphing_Mode7_new_physic.csv};
		
		\addplot[color=blue] table[x=t,y=f,col sep=comma]{./data/9cells_modes10_45/Morphing_Mode8_algeb.csv};
		\addplot[color=black] table[x=t,y=f,col sep=comma]{./data/9cells_modes10_45/Morphing_Mode8_new_physic.csv};
		
		\addplot[color=blue] table[x=t,y=f,col sep=comma]{./data/9cells_modes10_45/Morphing_Mode9_algeb.csv};
		\addplot[color=black] table[x=t,y=f,col sep=comma]{./data/9cells_modes10_45/Morphing_Mode9_new_physic.csv};
		
		\node at (axis cs:1,2.9421e9) [anchor=west, yshift = -0.4em] {{\scriptsize TM010}};
		\node at (axis cs:1,2.9457e9) [anchor=west] {{\scriptsize TM011}};
		\node at (axis cs:1,2.9562e9) [anchor=west] {{\scriptsize TM012}};
		\node at (axis cs:1,2.9738e9) [anchor=west] {{\scriptsize TM013}};
		\node at (axis cs:1,2.9983e9) [anchor=west] {{\scriptsize TM014}};
		\node at (axis cs:1,3.0294e9) [anchor=west] {{\scriptsize TM015}};
		\node at (axis cs:1,3.0670e9) [anchor=west] {{\scriptsize TM016}};
		\node at (axis cs:1,3.1109e9) [anchor=west] {{\scriptsize TM017}};
		\node at (axis cs:1,3.1608e9) [anchor=west] {{\scriptsize TM018}};

		\end{axis}
		\end{tikzpicture}%
}},draw,circle,minimum size=1cm] at (spypoint2) {};
\end{tikzpicture}
}
	\caption{Identification of the first monopole passband in the 9-cell TESLA cavity
	using algebraic and physical mappings from the TESLA cavity $(t = 0)$ to pillbox ($t = 1$).}
	\label{fig:9cells_modes1_9}
\end{figure}

The tracking plot for the first monopole passband (modes $j=1,\ldots,9$) can be seen in Fig.~\ref{fig:9cells_modes1_9} for both mappings, indicated in different colors.
We can observe that both the algebraic and the physical mapping yield the same classification in the pillbox cavity.

\begin{figure}
	\centering
	%!TEX root = ../main.tex
\begin{tikzpicture}
	\begin{axis}[width=\linewidth, height=0.75\linewidth,
		xlabel={t},ylabel={Frequency [Hz]}, 
		xlabel style = {yshift=.5em},
		legend pos=south east, 
		legend style={font=\footnotesize},
		xlabel={$t$}, 
		title style = {yshift=0.5em, xshift=0em},
		ylabel={$f ~ [\mathrm{Hz}]$ },
		ylabel style = {yshift=-1em},
		grid=both,
		minor grid style={gray!25},
		major grid style={gray!25},
		no marks, clip mode = individual]
	
		\legend{Modes 1 -- 9, Modes 10--27, Modes 28--45}
		
		\addplot[color=TUDa-3d]table[x=t,y=f,col sep=comma]{data/9cells_modes10_45/Morphing_Mode1_algeb.csv};
		\addplot[color=TUDa-7d] table[x=t,y=f,col sep=comma]{data/9cells_modes10_45/Morphing_Mode10_algeb.csv};
		\addplot[color=TUDa-9d] table[x=t,y=f,col sep=comma]{data/9cells_modes10_45/Morphing_Mode28_algeb.csv};
		
		\addplot[color=TUDa-3d] table[x=t,y=f,col sep=comma]{data/9cells_modes10_45/Morphing_Mode2_algeb.csv};
		\addplot[color=TUDa-3d] table[x=t,y=f,col sep=comma]{data/9cells_modes10_45/Morphing_Mode3_algeb.csv};
		\addplot[color=TUDa-3d] table[x=t,y=f,col sep=comma]{data/9cells_modes10_45/Morphing_Mode4_algeb.csv};
		\addplot[color=TUDa-3d] table[x=t,y=f,col sep=comma]{data/9cells_modes10_45/Morphing_Mode5_algeb.csv};
		\addplot[color=TUDa-3d] table[x=t,y=f,col sep=comma]{data/9cells_modes10_45/Morphing_Mode6_algeb.csv};
		\addplot[color=TUDa-3d] table[x=t,y=f,col sep=comma]{data/9cells_modes10_45/Morphing_Mode7_algeb.csv};
		\addplot[color=TUDa-3d] table[x=t,y=f,col sep=comma]{data/9cells_modes10_45/Morphing_Mode8_algeb.csv};
		\addplot[color=TUDa-3d] table[x=t,y=f,col sep=comma]{data/9cells_modes10_45/Morphing_Mode9_algeb.csv};
		
		\addplot[color=TUDa-7d] table[x=t,y=f,col sep=comma]{data/9cells_modes10_45/Morphing_Mode11_algeb.csv};
		\addplot[color=TUDa-7d] table[x=t,y=f,col sep=comma]{data/9cells_modes10_45/Morphing_Mode12_algeb.csv};
		\addplot[color=TUDa-7d] table[x=t,y=f,col sep=comma]{data/9cells_modes10_45/Morphing_Mode13_algeb.csv};
		\addplot[color=TUDa-7d] table[x=t,y=f,col sep=comma]{data/9cells_modes10_45/Morphing_Mode14_algeb.csv};
		\addplot[color=TUDa-7d] table[x=t,y=f,col sep=comma]{data/9cells_modes10_45/Morphing_Mode15_algeb.csv};
		\addplot[color=TUDa-7d] table[x=t,y=f,col sep=comma]{data/9cells_modes10_45/Morphing_Mode16_algeb.csv};
		\addplot[color=TUDa-7d] table[x=t,y=f,col sep=comma]{data/9cells_modes10_45/Morphing_Mode17_algeb.csv};
		\addplot[color=TUDa-7d] table[x=t,y=f,col sep=comma]{data/9cells_modes10_45/Morphing_Mode18_algeb.csv};
		\addplot[color=TUDa-7d] table[x=t,y=f,col sep=comma]{data/9cells_modes10_45/Morphing_Mode19_algeb.csv};
		\addplot[color=TUDa-7d] table[x=t,y=f,col sep=comma]{data/9cells_modes10_45/Morphing_Mode20_algeb.csv};
		\addplot[color=TUDa-7d] table[x=t,y=f,col sep=comma]{data/9cells_modes10_45/Morphing_Mode21_algeb.csv};
		\addplot[color=TUDa-7d] table[x=t,y=f,col sep=comma]{data/9cells_modes10_45/Morphing_Mode22_algeb.csv};
		\addplot[color=TUDa-7d] table[x=t,y=f,col sep=comma]{data/9cells_modes10_45/Morphing_Mode23_algeb.csv};
		\addplot[color=TUDa-7d] table[x=t,y=f,col sep=comma]{data/9cells_modes10_45/Morphing_Mode24_algeb.csv};
		\addplot[color=TUDa-7d] table[x=t,y=f,col sep=comma]{data/9cells_modes10_45/Morphing_Mode25_algeb.csv};
		\addplot[color=TUDa-7d] table[x=t,y=f,col sep=comma]{data/9cells_modes10_45/Morphing_Mode26_algeb.csv};
		\addplot[color=TUDa-7d] table[x=t,y=f,col sep=comma]{data/9cells_modes10_45/Morphing_Mode27_algeb.csv};
		
		\addplot[color=TUDa-9d] table[x=t,y=f,col sep=comma]{data/9cells_modes10_45/Morphing_Mode29_algeb.csv};
		\addplot[color=TUDa-9d] table[x=t,y=f,col sep=comma]{data/9cells_modes10_45/Morphing_Mode30_algeb.csv};
		\addplot[color=TUDa-9d] table[x=t,y=f,col sep=comma]{data/9cells_modes10_45/Morphing_Mode31_algeb.csv};
		\addplot[color=TUDa-9d] table[x=t,y=f,col sep=comma]{data/9cells_modes10_45/Morphing_Mode32_algeb.csv};
		\addplot[color=TUDa-9d] table[x=t,y=f,col sep=comma]{data/9cells_modes10_45/Morphing_Mode33_algeb.csv};
		\addplot[color=TUDa-9d] table[x=t,y=f,col sep=comma]{data/9cells_modes10_45/Morphing_Mode34_algeb.csv};
		\addplot[color=TUDa-9d] table[x=t,y=f,col sep=comma]{data/9cells_modes10_45/Morphing_Mode35_algeb.csv};
		\addplot[color=TUDa-9d] table[x=t,y=f,col sep=comma]{data/9cells_modes10_45/Morphing_Mode36_algeb.csv};
		\addplot[color=TUDa-9d] table[x=t,y=f,col sep=comma]{data/9cells_modes10_45/Morphing_Mode37_algeb.csv};
		\addplot[color=TUDa-9d] table[x=t,y=f,col sep=comma]{data/9cells_modes10_45/Morphing_Mode38_algeb.csv};
		\addplot[color=TUDa-9d] table[x=t,y=f,col sep=comma]{data/9cells_modes10_45/Morphing_Mode39_algeb.csv};
		\addplot[color=TUDa-9d] table[x=t,y=f,col sep=comma]{data/9cells_modes10_45/Morphing_Mode40_algeb.csv};
		\addplot[color=TUDa-9d] table[x=t,y=f,col sep=comma]{data/9cells_modes10_45/Morphing_Mode41_algeb.csv};
		\addplot[color=TUDa-9d] table[x=t,y=f,col sep=comma]{data/9cells_modes10_45/Morphing_Mode42_algeb.csv};
		\addplot[color=TUDa-9d] table[x=t,y=f,col sep=comma]{data/9cells_modes10_45/Morphing_Mode43_algeb.csv};
		\addplot[color=TUDa-9d] table[x=t,y=f,col sep=comma]{data/9cells_modes10_45/Morphing_Mode44_algeb.csv};
		\addplot[color=TUDa-9d] table[x=t,y=f,col sep=comma]{data/9cells_modes10_45/Morphing_Mode45_algeb.csv};
	\end{axis}
\end{tikzpicture}
	\caption{Identification of eigenmodes $j=1,\ldots,45$ of the 9-cell TESLA cavity applying the algebraic mapping from TESLA cavity ($t = 0$) to pillbox ($t = 1$).}
	\label{fig:results9cellsmodes1_45}
\end{figure}
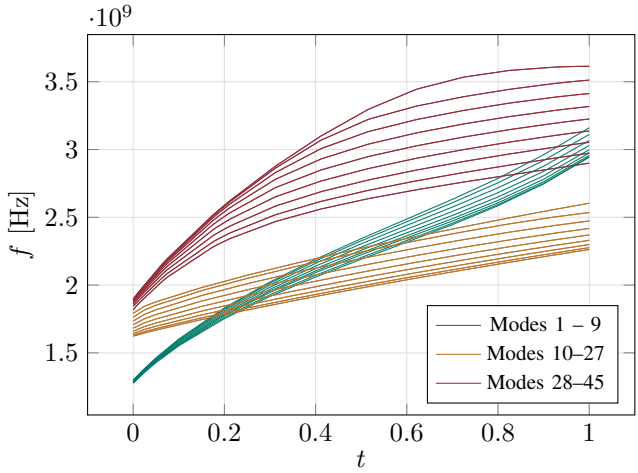

Fig.~\ref{fig:results9cellsmodes1_45} visualizes the tracking results for the first 45 eigenmodes in the TESLA cavity.
We can see that the tracking yields smooth results.
The crossings of the eigenvalues and the overlapping of eigenmode passbands emphasize the need for a tracking procedure to clearly match and classify the eigenmodes along the shape morphing.

\begin{figure}
	\centering
	\begin{tikzpicture}
\begin{axis}[width=0.75\linewidth, height=0.75\linewidth,
	xlabel={t},ylabel={Frequency [Hz]}, 
	xlabel style = {yshift=.5em},
	legend pos=south east, 
	xmin = -0.05,
	xmax = 1.18,
	xtick = {0, 0.2, 0.4, 0.6, 0.8, 1.0},
	legend style={font=\footnotesize},
	xlabel={$t$}, 
	title style = {yshift=-.5em, xshift=4em},
	ylabel={$f ~ [\mathrm{Hz}]$ },
	ylabel style = {yshift=-1em},
	grid=both,
	minor grid style={gray!25},
	major grid style={gray!25},
	no marks, clip mode = individual]

\addplot[only marks, red, mark=x, mark size= 1pt] coordinates {				
    (1,2.294e9) 
	(1,2.412e9)
	(1,2.598e9)
	(1,2.837e9)
	(1,3.118e9)
	(1,3.430e9)
	(1,2.942e9)
	(1,2.974e9)
	(1,3.066e9)};
	
\node at (axis cs:1,2.294e9) [anchor=west] {{\scriptsize TE111}};
\node at (axis cs:1,2.412e9) [anchor=west] {{\scriptsize TE112}};
\node at (axis cs:1,2.598e9) [anchor=west] {{\scriptsize TE113}};
\node at (axis cs:1,2.837e9) [anchor=west] {{\scriptsize TE114}};
\node at (axis cs:1,3.118e9) [anchor=west] {{\scriptsize TE115}};
\node at (axis cs:1,3.430e9) [anchor=west] {{\scriptsize TE116}};
\node at (axis cs:1,2.942e9) [anchor=west] {{\scriptsize TM010}};
\node at (axis cs:1,2.974e9) [anchor=west] {{\scriptsize TM011}};
\node at (axis cs:1,3.066e9) [anchor=west] {{\scriptsize TM012}};

\addplot[color=blue] table[x=t,y=f,col sep=comma]%
{data/3cells_modes1_15_algebraicTracking/Morphing_Mode1_algeb.csv};

\addplot[color=blue] table[x=t,y=f,col sep=comma]{data/3cells_modes1_15_algebraicTracking/Morphing_Mode2_algeb.csv};
\addplot[color=blue] table[x=t,y=f,col sep=comma]{data/3cells_modes1_15_algebraicTracking/Morphing_Mode3_algeb.csv};
\addplot[color=blue] table[x=t,y=f,col sep=comma]{data/3cells_modes1_15_algebraicTracking/Morphing_Mode4_algeb.csv};
\addplot[color=blue] table[x=t,y=f,col sep=comma]{data/3cells_modes1_15_algebraicTracking/Morphing_Mode5_algeb.csv};
\addplot[color=blue] table[x=t,y=f,col sep=comma]{data/3cells_modes1_15_algebraicTracking/Morphing_Mode6_algeb.csv};
\addplot[color=blue] table[x=t,y=f,col sep=comma]{data/3cells_modes1_15_algebraicTracking/Morphing_Mode7_algeb.csv};
\addplot[color=blue] table[x=t,y=f,col sep=comma]{data/3cells_modes1_15_algebraicTracking/Morphing_Mode8_algeb.csv};
\addplot[color=blue] table[x=t,y=f,col sep=comma]{data/3cells_modes1_15_algebraicTracking/Morphing_Mode9_algeb.csv};
\addplot[color=blue] table[x=t,y=f,col sep=comma]{data/3cells_modes1_15_algebraicTracking/Morphing_Mode10_algeb.csv};
\addplot[color=blue] table[x=t,y=f,col sep=comma]{data/3cells_modes1_15_algebraicTracking/Morphing_Mode11_algeb.csv};
\addplot[color=blue] table[x=t,y=f,col sep=comma]{data/3cells_modes1_15_algebraicTracking/Morphing_Mode12_algeb.csv};
\addplot[color=blue] table[x=t,y=f,col sep=comma]{data/3cells_modes1_15_algebraicTracking/Morphing_Mode13_algeb.csv};
\addplot[color=blue] table[x=t,y=f,col sep=comma]{data/3cells_modes1_15_algebraicTracking/Morphing_Mode14_algeb.csv};
\addplot[color=blue] table[x=t,y=f,col sep=comma]{data/3cells_modes1_15_algebraicTracking/Morphing_Mode15_algeb.csv};

\coordinate (spypoint2) at (axis cs:1,2.975e9);
	\end{axis}
	
\node[pin={[pin distance=0.6cm]-10:{%
		\begin{tikzpicture}[tight background,background rectangle/.style={fill=gray!10}, show background rectangle, framed]
		\begin{axis}[
		no markers,
		grid=major,
		tiny, clip mode = individual,
		xmin=0.99, xmax=1.02, width=.35\linewidth, height=.5\linewidth, xtick = {0.99, 1.0}%, 1.01, 1.02}
		, ytick = {2.8e9, 2.9e9, 3.0e9, 3.1e9, 3.2e9}, ymin = 2.8e9, ymax=3.2e9
		]
		\addplot[only marks, red, mark=x, mark size= 2pt] coordinates {(1,2.9421e9) 
	(1,2.9421e9)
	(1,2.9735e9)
	(1,3.0659e9)
	(1,2.8372e9)
	(1,3.1180e9)};
	\addplot[color=blue] table[x=t,y=f,col sep=comma]%
{data/3cells_modes1_15_algebraicTracking/Morphing_Mode1_algeb.csv};

\addplot[color=blue] table[x=t,y=f,col sep=comma]{data/3cells_modes1_15_algebraicTracking/Morphing_Mode2_algeb.csv};
\addplot[color=blue] table[x=t,y=f,col sep=comma]{data/3cells_modes1_15_algebraicTracking/Morphing_Mode3_algeb.csv};
\addplot[color=blue] table[x=t,y=f,col sep=comma]{data/3cells_modes1_15_algebraicTracking/Morphing_Mode4_algeb.csv};
\addplot[color=blue] table[x=t,y=f,col sep=comma]{data/3cells_modes1_15_algebraicTracking/Morphing_Mode5_algeb.csv};
\addplot[color=blue] table[x=t,y=f,col sep=comma]{data/3cells_modes1_15_algebraicTracking/Morphing_Mode6_algeb.csv};
\addplot[color=blue] table[x=t,y=f,col sep=comma]{data/3cells_modes1_15_algebraicTracking/Morphing_Mode7_algeb.csv};
\addplot[color=blue] table[x=t,y=f,col sep=comma]{data/3cells_modes1_15_algebraicTracking/Morphing_Mode8_algeb.csv};
\addplot[color=blue] table[x=t,y=f,col sep=comma]{data/3cells_modes1_15_algebraicTracking/Morphing_Mode9_algeb.csv};
\addplot[color=blue] table[x=t,y=f,col sep=comma]{data/3cells_modes1_15_algebraicTracking/Morphing_Mode10_algeb.csv};
\addplot[color=blue] table[x=t,y=f,col sep=comma]{data/3cells_modes1_15_algebraicTracking/Morphing_Mode11_algeb.csv};
\addplot[color=blue] table[x=t,y=f,col sep=comma]{data/3cells_modes1_15_algebraicTracking/Morphing_Mode12_algeb.csv};
\addplot[color=blue] table[x=t,y=f,col sep=comma]{data/3cells_modes1_15_algebraicTracking/Morphing_Mode13_algeb.csv};
\addplot[color=blue] table[x=t,y=f,col sep=comma]{data/3cells_modes1_15_algebraicTracking/Morphing_Mode14_algeb.csv};
\addplot[color=blue] table[x=t,y=f,col sep=comma]{data/3cells_modes1_15_algebraicTracking/Morphing_Mode15_algeb.csv};
		
\node at (axis cs:1,2.8372e9) [anchor=west] {{\scriptsize TE114}};
\node at (axis cs:1,2.9421e9) [anchor=west] {{\scriptsize TM010}};
\node at (axis cs:1,2.9735e9) [anchor=west] {{\scriptsize TM011}};
\node at (axis cs:1,3.0659e9) [anchor=west] {{\scriptsize TM012}};
\node at (axis cs:1,3.1180e9) [anchor=west] {{\scriptsize TE115}};

\end{axis}
\end{tikzpicture}%
}},draw,circle,minimum size=1cm] at (spypoint2) {};	
\end{tikzpicture}
	\caption{Identification of the eigenmodes $j=1,\ldots,15$ %monopole and first and second dipole passband 
		of the 3-cell TESLA cavity applying the algebraic mapping from TESLA cavity ($t=0$) to pillbox ($t=1$).}
	\label{fig:3cells_modes1_15}
\end{figure}
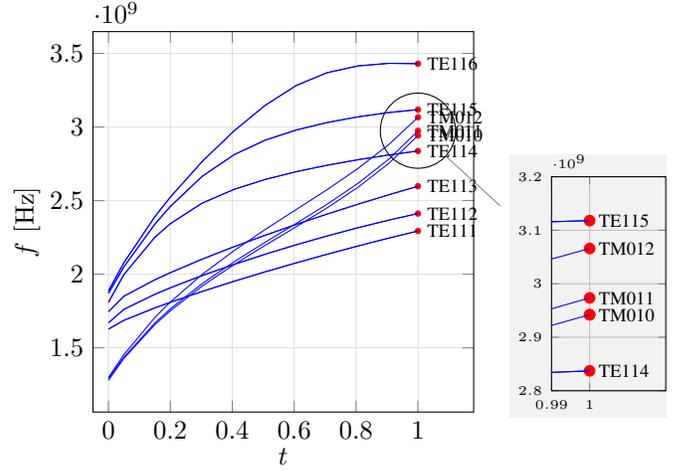

Our eigenmode classification method is not limited to the 9-cell TESLA cavity, but can also be applied to a design with fewer cells. 
Fig.~\ref{fig:3cells_modes1_15} demonstrates the tracking for a 3-cell TESLA cavity showing only the algebraic mapping for clarity, classifying the first 15 eigenmodes, corresponding to the first monopole and the first and second dipole passband.
We observe analogous results to those of the 9-cell geometry.
Each passband consists of the same number of eigenmodes as the cavity of cells, if disregarding the multiplicity of eigenmodes.
In the first monopole passband, the lowest eigenmode is classified as $\mathrm{TM}0\myspace1\myspace0$ and in the first dipole passband as $\mathrm{TE}1\myspace1\myspace1$.
The longitudinal index is accordingly increased.

%---------------------------------------------------------1---------------
\subsection{Exploration of Eigenmodes}
\begin{figure}
	\centering
	\begin{tikzpicture}
\begin{axis}[width=\linewidth, height=0.75\linewidth,
	xlabel={t},ylabel={Frequency [Hz]}, 
	xlabel style = {yshift=.5em},
	xmax = 1.18,
	xtick = {0, 0.2, 0.4, 0.6, 0.8, 1.0},
	legend pos=south east, 
	legend style={font=\footnotesize},
	xlabel={$t$}, 
	title style = {yshift=-.5em, xshift=4em},
	ylabel={$f ~ [\mathrm{Hz}]$ },
	ylabel style = {yshift=-1em},
	grid=both,
	minor grid style={gray!25},
	major grid style={gray!25},
	no marks, clip mode = individual]

\addplot[color=blue] table[x=t,y=f,col sep=comma]{data/Morphing_Mode66_algeb.csv};
\addplot[color=black] table[x=t,y=f,col sep=comma]{data/Morphing_Mode66_physic.csv};
\legend{Algebraic,Physical}
\addplot[only marks, red, mark=x, mark size= 1pt] coordinates {(1,3.83e9)};
\node at (axis cs:1,3.83e9) [anchor=west] {{\scriptsize TM0117}};
\end{axis}
\end{tikzpicture}
	\caption{Exploration of eigenmode $\mathrm{TM}0\myspace1\myspace17$ from the pillbox ($t = 1$) down to the 9-cell TESLA cavity ($t = 0$).}
	\label{fig:backwardTracking}
\end{figure}
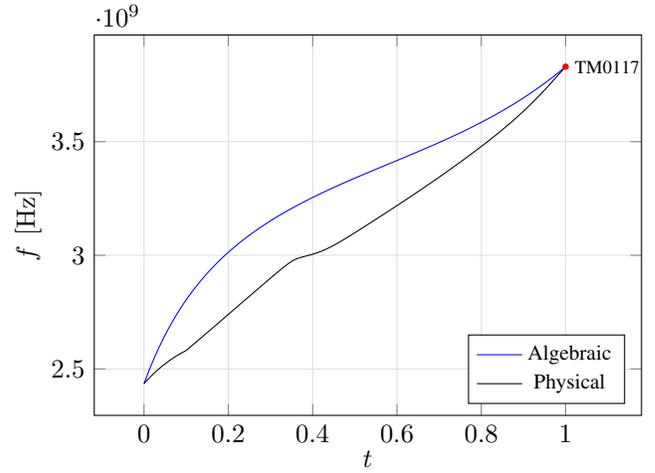

\begin{figure}
    \begin{subfigure}[b]{1.0\linewidth}
		\centering
		\includegraphics[width=1.0\linewidth, trim =30mm 72mm 30mm 72mm, clip]{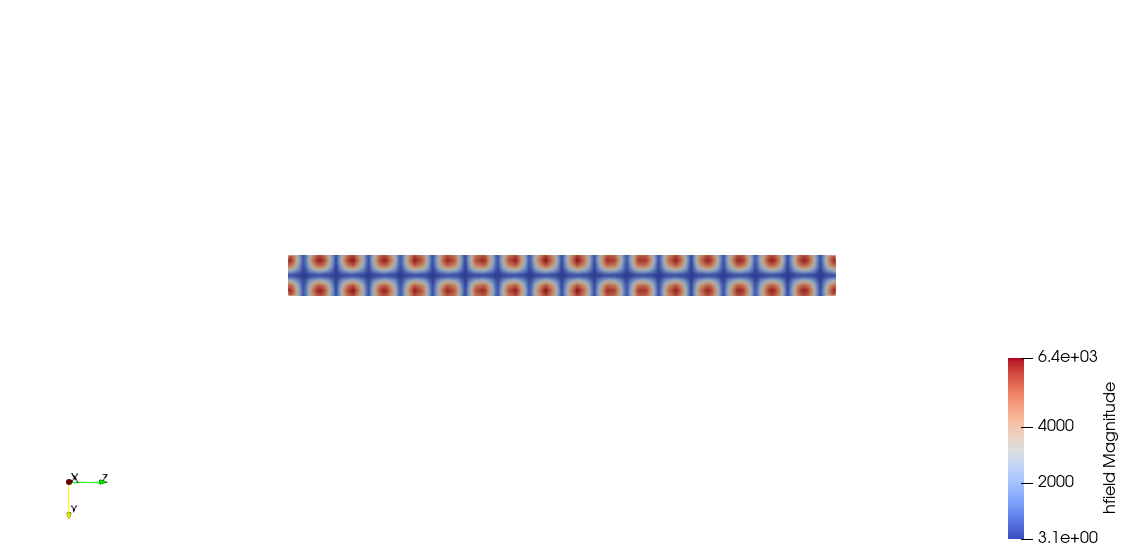}
		\caption{Magnitude of the magnetic field at $t=0.00$}
		\label{fig:TM0117pillbox}
	\end{subfigure}
	\begin{subfigure}[b]{1.0\linewidth}
		\centering
		\includegraphics[width=1.0\linewidth, trim =30mm 72mm 30mm 72mm, clip]{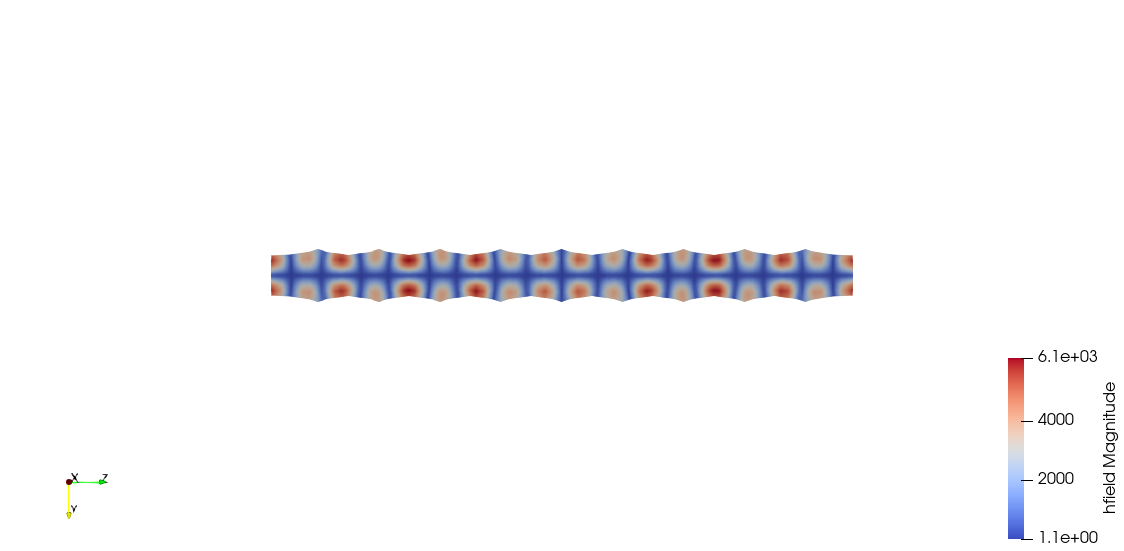}
		\caption{Magnitude of the magnetic field at $t=0.15$}
% 		\label{fig:TM0117middle}
	\end{subfigure}
	\begin{subfigure}[b]{1.0\linewidth}
		\centering
		\includegraphics[width=1.0\linewidth, trim =30mm 72mm 30mm 72mm, clip]{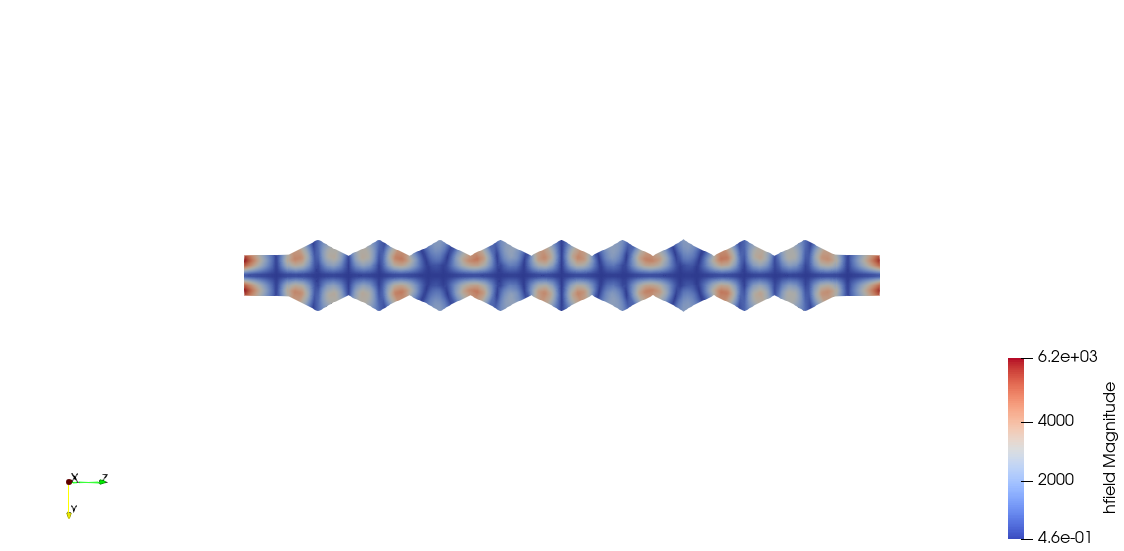}
		\caption{Magnitude of the magnetic field at $t=0.40$}
	\end{subfigure}
	\begin{subfigure}[b]{1.0\linewidth}
		\centering
		\includegraphics[width=1.0\linewidth, trim =30mm 72mm 30mm 72mm, clip]{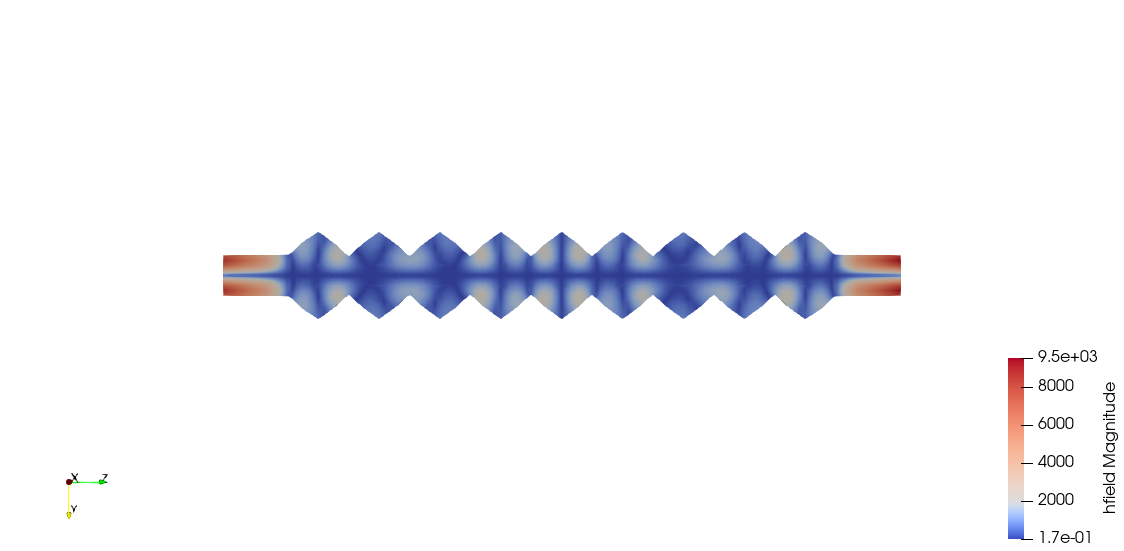}
		\caption{Magnitude of the magnetic field at $t=0.60$}
	\end{subfigure}
	\begin{subfigure}[b]{1.0\linewidth}
		\centering
		\includegraphics[width=1.0\linewidth, trim =30mm 72mm 30mm 72mm, clip]{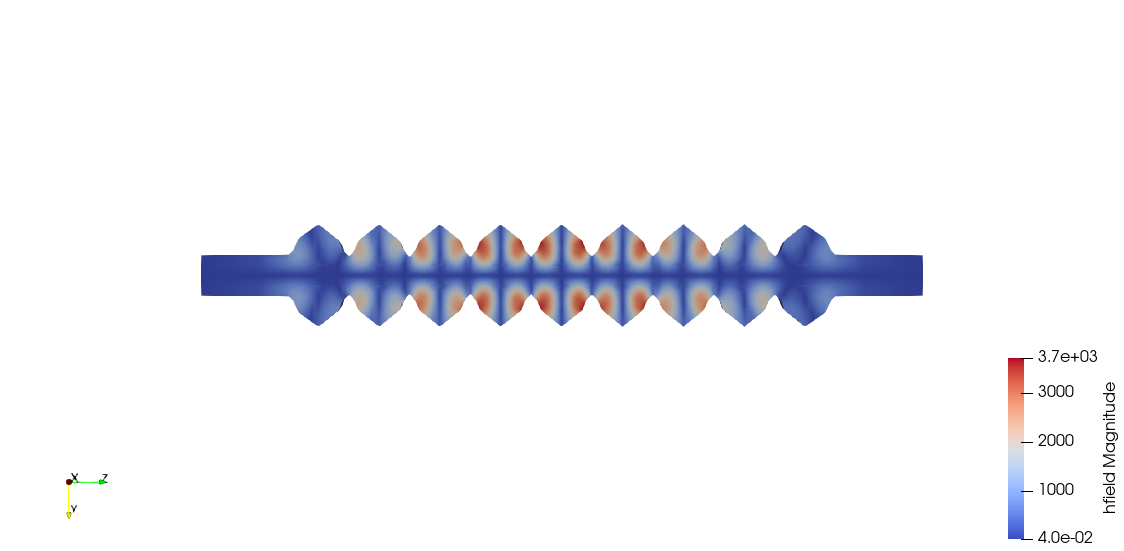}
		\caption{Magnitude of the magnetic field at $t=0.80$}
	\end{subfigure}
	\begin{subfigure}[b]{1.0\linewidth}
		\centering
		\includegraphics[width=1.0\linewidth, trim =30mm 72mm 30mm 72mm, clip]{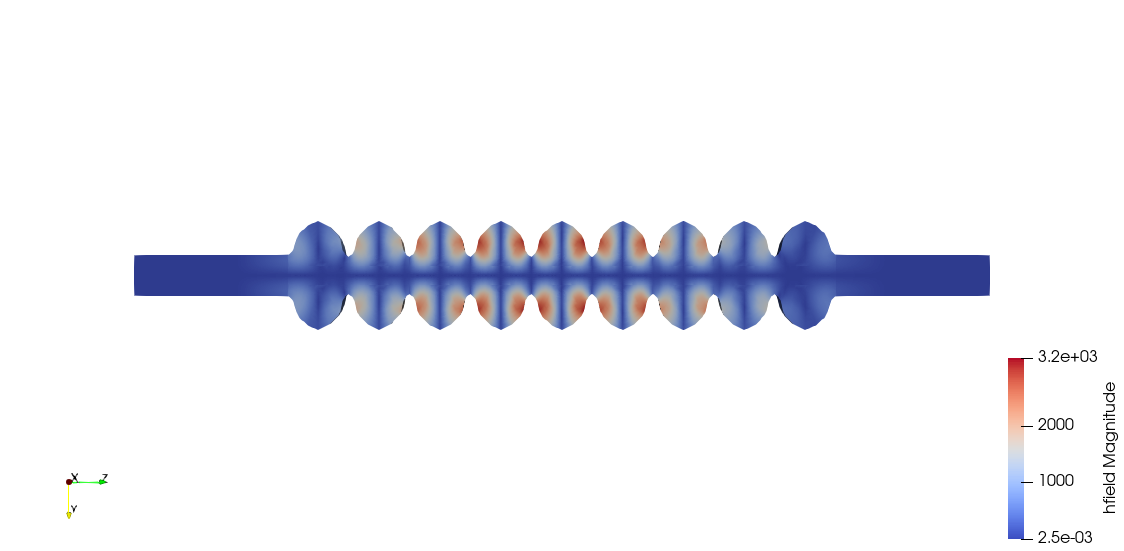}
		\caption{Magnitude of the magnetic field at $t=0.95$}
	\end{subfigure}
	\begin{subfigure}[b]{1.0\linewidth}
		\centering
		\includegraphics[width=1.0\linewidth, trim =30mm 72mm 30mm 72mm, clip]{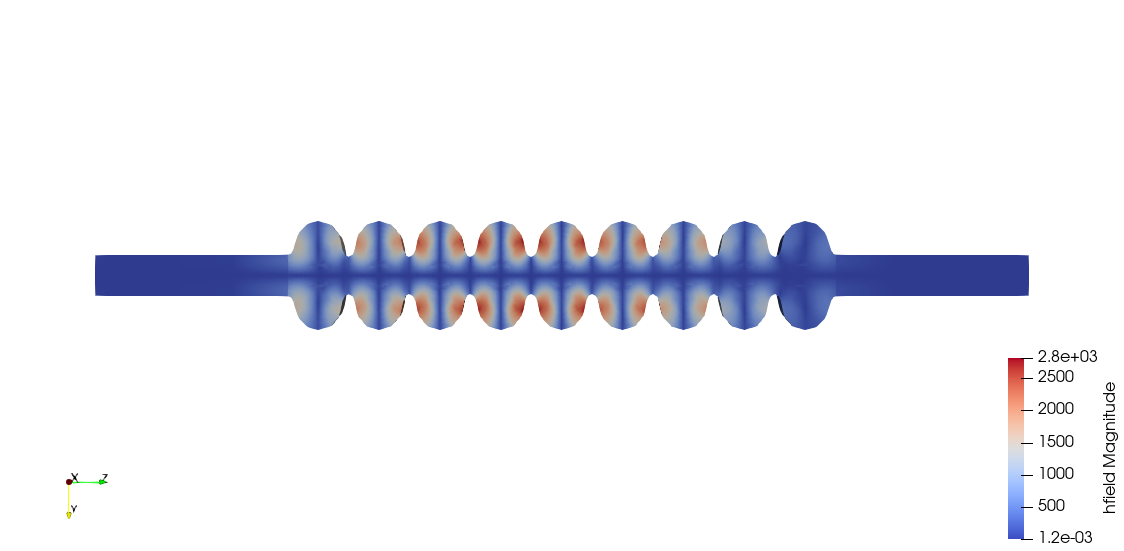}
		\caption{Magnitude of the magnetic field at $t=1.00$}
		\label{fig:TM0117tesla}
	\end{subfigure}
	\caption{Magnetic field magnitude of the eigenmode $\mathrm{TM}0\myspace1\myspace17$ 
	(i.e. the highest eigenmode of the second monopole passband) along the physical deformation for exploration in the TESLA cavity.}
	\label{fig:TM0117}
\end{figure}

%------------------------------------------------------------------------
The described tracking algorithm can be used to directly calculate a specific eigenmode in the TESLA cavity without the need to identify all eigenmodes.
For this purpose, the eigenvalue $\lambda_{1,j}$ for chosen indices $m$, $n$ and $p$ is determined for the pillbox cavity and the corresponding eigenpair is tracked as described in Section~\ref{sec:backwardTracking}. This is exemplified for eigenmode $\mathrm{TM}0\myspace1\myspace17$, which is the highest eigenmode in the second monopole passband for the 9-cell TESLA cavity in Fig.~\ref{fig:backwardTracking}.
Firstly, applying~\eqref{eq:resonanceFrequenciesTM}, the resonance frequency for the indices~$m=0$, $n = 1$ and $p = 17$ is computed at approx. \SI{3.83}{\giga \hertz}.
The eigenvalue problem~\eqref{eq:eigprob} is solved for the pillbox cavity and the eigenmode with the closest eigenfrequency is tracked to approx. \SI{2.44}{\giga \hertz} in the TESLA cavity.
The tracking results are displayed in Fig.~\ref{fig:backwardTracking}. 
The results for the physical mapping are plotted and in Fig.~\ref{fig:TM0117}, where the deformation of the magnitude of the magnetic field is illustrated.

%------------------------------------------------------------------------
\subsection{Efficiency}
%------------------------------------------------------------------------
If only one mode at a time is considered, then using Newton's method to solve the eigenvalue problem in each step, see e.g. \cite{Georg_2019aa} may be an efficient alternative to the Arnoldi method used above. However, this might be less robust since Newton's method may converge to a wrong mode.
In particular, when the step size $h_i$ is small, one may expect fast convergence of Newton's method due to availability of good initial guesses from the previous step, see \eqref{eq:prediction}. Let us consider an example in which we use the number of solved linear systems and the time for tracking the accelerating eigenmode for the \mbox{9-cell} cavity as efficiency measure. We employ the algebraic mapping for a fixed stepsize of $h = 0.1$ such that no step size reduction was necessary. 
The computations are performed on a workstation with an Intel(R) CPU i7-3820 3.6~GHz processor and 13~GB RAM.
The cost for assembly of the system matrices $\mathbf{K}(0)$, $\mathbf{M}(0)$, $\mathbf{K}(1)$ and $\mathbf{M}(1)$ is excluded.
All time measurements are repeated 10 times and averaged.
The eigenvalue problems are solved using Matlab's \texttt{eigs} which uses internally ARPACK \cite{Mathworks_2020aa,Lehoucq_1998aa}. Classifying the accelerating eigenmode using the presented method requires 1571 systems of equations to be solved and on average \SI{58.69}{\s}, while the tracking based on Newton's method solves 297 linear systems in \SI{37.29}{\s}.

If $N_\textrm{tracked}>1$ eigenmodes are of interest, then the efficiency of the tracking method in section~\ref{sec:tracking} can be increased by considering all of them at once. Since we always compute $N>N_\textrm{tracked}$ eigenpairs per step, this does not significantly increase the computational time of the eigenvalue solver, which is the most expensive component of the procedure. Yet, the complexity of the matching increases quadratically, or more precisely $N_\textrm{tracked}\cdot N$, since each mode of interest must be compared with all modes at the next step. Nevertheless, those costs remain negligible in comparison to the eigenvalue solver, even if hundreds of modes would be tracked.

%------------------------------------------------------------------------
\section{Conclusion and Outlook} \label{sec:conclusion}
%------------------------------------------------------------------------
In this work, we investigate an eigenmode tracking procedure and apply it for the automatic classification of eigenmodes in electromagnetic cavities. 
To this end, we employ shape morphing from complex geometries to the analytical pillbox cavity.
By matching our numerical results with the analytic values, we are able to convey the well-known nomenclature based on the type of the eigenmodes (TE or TM) and the directional indices $m$, $n$ and $p$ and their intuition on more complex cavity shapes.
By performing our algorithm in a backward manner, we can explore specific eigenmodes in a cavity. 

We have shown that despite of eigenvalue crossings, our approach reliably tracks the eigenmodes also in 9-cell TESLA cavities, in contrast to field extrema counting methods, e.g.~\cite{Brackebusch_2014aa}.
The numerical efficiency of our tracking method could be further improved by enhancing the prediction step~\eqref{eq:prediction} with higher order Taylor expansion~\cite{Jorkowski_2018aa} or by reducing our system matrices to a subspace of eigenpairs~\cite{Yang_2007ab}.
Furthermore, the accuracy of the derivatives of the system matrices can be improved by formulating them as shape derivatives along the geometry deformation~\cite{Sturm_2015ab}.


\end{document}